\documentclass[sigconf]{acmart}
\usepackage{amsmath}
\usepackage[noend]{algorithmic}
\usepackage[nameinlink,capitalize]{cleveref}
\usepackage{float}
\usepackage{algorithm}
\usepackage{subcaption}
\usepackage{dsfont}
\usepackage{multirow}
\usepackage{array}
\usepackage{enumitem}
\usepackage{multibib}
\newcites{AP}{Appendix References}
\usepackage[colorinlistoftodos,textsize=tiny]{todonotes}
\usepackage{balance}

\usepackage{soul}
\definecolor{navy}{rgb}{0.1, 0.1, 0.8}
\definecolor[named]{gray}{rgb}{0.4, 0.4, 0.4}
\definecolor[named]{olive}{rgb}{0.1, 0.5, 0.1}
\definecolor[named]{ruby}{rgb}{0.8, 0.1, 0.3}
\definecolor{darkpastelgreen}{rgb}{0.01, 0.75, 0.24}
\definecolor{celestialblue}{rgb}{0.29, 0.59, 0.82}
\definecolor{coral}{rgb}{1.0, 0.5, 0.31}
\definecolor{Goldenrod}{rgb}{0.8,0.8,0}

\newcommand{\eat}[1]{}
\newcommand{\rev}[1]{{#1}}

\newcommand{\revA}[1]{{#1}}
\newcommand{\verify}[1]{{}}
\newcommand{\verifyK}[1]{{#1}}
\newcommand{\NOTE}[2]{ }
\newcommand{\nb}[1]{}

\DeclareMathOperator{\E}{\mathbb{E}}
\DeclareMathOperator{\Prob}{\mathbb{P}}
\DeclareMathOperator{\Real}{\mathbb{R}}

\DeclareMathOperator{\His}{\mathcal{H}}
\newcommand{\bracket}[1]{\left[#1\right]}

\newcommand{\numberthis}{\addtocounter{equation}{1}\tag{\theequation}}

\def\BibTeX{{\rm B\kern-.05em{\sc i\kern-.025em b}\kern-.08emT\kern-.1667em\lower.7ex\hbox{E}\kern-.125emX}}

\newcolumntype{L}[1]{>{\raggedright\let\newline\\\arraybackslash\hspace{0pt}}m{#1}}
\newcolumntype{C}[1]{>{\centering\let\newline\\\arraybackslash\hspace{0pt}}m{#1}}
\newcolumntype{R}[1]{>{\raggedleft\let\newline\\\arraybackslash\hspace{0pt}}m{#1}}
\copyrightyear{2020}
\acmYear{2020}
\setcopyright{acmcopyright}
\acmConference[WSDM '20]{The Thirteenth ACM International Conference on Web Search and Data Mining}{February 3--7, 2020}{Houston, TX, USA}
\acmBooktitle{The Thirteenth ACM International Conference on Web Search and Data Mining (WSDM '20), February 3--7, 2020, Houston, TX, USA}
\acmPrice{15.00}
\acmDOI{10.1145/3336191.3371821}
\acmISBN{978-1-4503-6822-3/20/02}

\begin{document}
\fancyhead{}
 \newcommand{\titlename}{\verifyK{Modeling Information Cascades with Self-exciting Processes via Generalized Epidemic Models}}
\title{\titlename}

\author{Quyu Kong}
\affiliation{%
  \institution{Australian National University \&\\ UTS \& Data61, CSIRO}
  \city{Canberra}
  \country{Australia}}
\email{quyu.kong@anu.edu.au}

\author{Marian-Andrei Rizoiu}
\affiliation{%
  \institution{University of Technology Sydney \& Data61, CSIRO}
  \city{Sydney}
  \country{Australia}
}
\email{marian-andrei.rizoiu@uts.edu.au}

\author{Lexing Xie}
\affiliation{%
 \institution{Australian National University \& Data61, CSIRO}
 \city{Canberra}
  \country{Australia}}
  \email{lexing.xie@anu.edu.au}

\begin{abstract}
	Epidemic models and self-exciting processes are two types of models used to describe diffusion phenomena online and offline.
    These models were originally developed in different scientific communities, and their commonalities are under-explored.
    This work establishes, for the first time, a general connection between the two model classes via three new mathematical components. 
    The first is a generalized version of stochastic Susceptible-Infected-Recovered (SIR) model with arbitrary recovery time distributions; 
    the second is the relationship between the (latent and arbitrary) recovery time distribution, recovery hazard function, and the infection kernel of self-exciting processes; 
    the third includes methods for simulating, fitting, evaluating and predicting the generalized process. 
    On three large Twitter diffusion datasets, we conduct goodness-of-fit tests and holdout log-likelihood evaluation of self-exciting processes with three infection kernels --- exponential, power-law and Tsallis Q-exponential. 
    We show that the modeling performance of the infection kernels varies with respect to the temporal structures of diffusions, and also with respect to user behavior, such as the likelihood of being bots.
    We further improve the prediction of popularity by combining two models that are identified as complementary by the goodness-of-fit tests.
\end{abstract}

\keywords{Information Diffusion, Hawkes Processes, Epidemic Models}

\maketitle

\section{Introduction}

Epidemic models and self-exciting processes are two classes of mathematical models that have evolved separately and been applied in distinct problem domains, one in epidemiology~\citep{kermack1927contribution} and the other in seismology~\citep{hawkes1971spectra,ogata1978asymptotic}, finance~\citep{bacry2015hawkes}, \revA{and} neural science~\citep{johnson1996point}.
\emph{Epidemic models} typically divide the population into compartments, such as Susceptible, Infected and Recovered for the Susceptible-Infected-Recovered (SIR) model, and describe the transitions between compartments as deterministic or stochastic processes. \emph{Self-exciting point processes} are a class of processes in which the occurrence of each event increases the likelihood of future events using time-decaying kernel functions.
Both models have been used to describe events in \revA{the} physical world, as well as online information diffusions~\citep{zhao2013sir,martin2016exploring,zarezade2017redqueen,li2017learning}. 
This paper aims to establish a mathematical connection between these two model classes. 
By \revA{achieving this, this work contributes}: 
1)~new expressive models for self-exciting processes in finite populations; 
2)~methods that account for unobserved recovery events, which are common in real-world epidemiological data; 
3)~new tools and insights into online information diffusion.

The Hawkes process with exponential kernels and stochastic SIR process have been recently shown~\citep{Rizoiu2017c} to share a connection \revA{via the} infection intensity function when the recovery time in the SIR model is \emph{latent}.
However, this result is restricted to one particular parametric family of self-exciting processes, whereas Hawkes processes allow a richer set of kernel functions, and an inequality of the connection has been overlooked.
These observations lead to the question: 
\textbf{How to both broaden and deepen the connection between epidemic models and Hawkes processes?}
The broadening is with respect to arbitrary recovery time distributions and kernel functions, while the deepening is with respect to the mathematical relationships between two model classes.
\revA{To address these}, we \revA{propose a} generalized stochastic SIR process \revA{in which} infected individuals recover independently \revA{following an} arbitrary distribution of recovery times. 
\revA{Next, we link this process to a} finite-population Hawkes process (dubbed  \emph{HawkesN}~\cite{Rizoiu2017c}) \revA{by showing that}
the Complementary Cumulative Distribution Function (CCDF) of the recovery time (in SIR), given the infection event history, is an upper bound of the \revA{HawkesN} kernel.
We derive relationships among three key functions: the kernel function in HawkesN, the SIR recovery time distribution, and the recovery hazard function.
We empirically evaluate the accuracy of recovering original parameters of stochastic 
SIR models from a HawkesN model fitted only on infection events.

Connecting \revA{the two} model classes will enrich the computational tools \revA{of} both. 
One challenge emerges --- \textbf{what tools can be developed \revA{and applied through the generalized connection to both classes of models?}}
We first enrich the generalized SIR with concepts from Hawkes processes including \emph{event marks} (features associated with events) and \emph{branching factors} (expected number of future events generated by a new event). We then show a simulation algorithm for the generalized SIR process by paired-sampling of infection and recovery times.
We also present maximum log-likelihood procedures for estimating the parameters, 2 metrics for measuring goodness-of-fit and approaches for predicting final diffusion popularity, \revA{for} SIR and HawkesN processes with general kernels.

While generalized models allow flexibility in the choice of parametric forms, it is important to understand \textbf{how the performances of different model formulations vary on diffusions?} 
On three large Twitter diffusion datasets, we show that the HawkesN model with different kernels demonstrates diverse modeling capability on diffusions with distinct temporal dynamics.
For instance, on one of the datasets, \textit{NEWS}, the HawkesN model with an exponential kernel tends to fit diffusions that are larger in event counts and shorter in time frames. We show that this can result from \revA{the participation of automated} bots \revA{to the online} diffusions.
These observations lead us to combine models for predicting diffusion final popularity, which outperforms all other models.

The main contributions of this work include:
\begin{itemize}[leftmargin=*]
    \item A generalized stochastic SIR process with arbitrary recovery time distributions and their connection to HawkesN processes with monotonically time-decaying kernels. The generalized model is equipped with concepts from Hawkes processes including event marks and branching factors.
    \item A set of tools including simulation, parameter estimation, evaluation and popularity prediction algorithms for SIR processes with general recovery time distributions.
    \item A series of fitting, model comparison and prediction results on real-world Twitter diffusion data. 
    We observe that the performances of general SIR processes with different recovery distributions vary with respect to diffusion dynamics. In prediction experiments, a combined model performs the best. 
\end{itemize}

\noindent{\bf Related work. } Effort has been put into generalizing epidemic models. \citet{keeling1997disease} reformulate the deterministic epidemic model as integro-differential equations, and impose a Gaussian distribution on the recovery times. \citet{streftaris2012non} specify the recovery times following a Weibull distribution and \citet{Routledge2018} model them using a Rayleigh distribution. On the Hawkes processes front, a rich set of kernel functions are available including power-law~\citep{Mishra2016FeaturePrediction}, piece-wise linear~\citep{zhou2013learning}, Tsallis Q-Exponential~\citep{lima2018hawkes}, and general function approximators such as neural networks~\citep{jing2017neural,Mishra2018ModelingPopularity,du2016recurrent,mei2017neural}. 
\revA{Our} work links the developments from both model classes \revA{via the proposed} generalized connection.

In terms of the study of information diffusion \revA{using} epidemic models, \citet{kimura2009efficient} first apply the SIS model, which allows nodes to be activated multiple times, to study information diffusion in a network. \citet{jin2013epidemiological} use an enhanced SEIZ, which introduces an extra \textit{Exposed} state~(E) to the SIR model for capturing a incubation period, to detect rumors from Twitter cascades. 
\revA{When studying online diffusion using self-exciting processes,}
\citet{Zhao2015SEISMIC:Popularity} and \citet{Mishra2016FeaturePrediction} both employ power-law kernel functions with Hawkes processes, which achieve state-of-art performance in popularity prediction.
\citet{Rizoiu2017c} apply HawkesN with an exponential kernel that outperforms the Hawkes counterpart in terms of holdout log-likelihood values.
Different from these works which show superior performance for a specific form in one or two evaluation tasks, our analysis \revA{corroborates} several aspects of tests including goodness-of-fit, holdout log-likelihood and \verifyK{prediction}.

\begin{figure}[!tbp]
	\centering
	\includegraphics[width=0.4\textwidth]{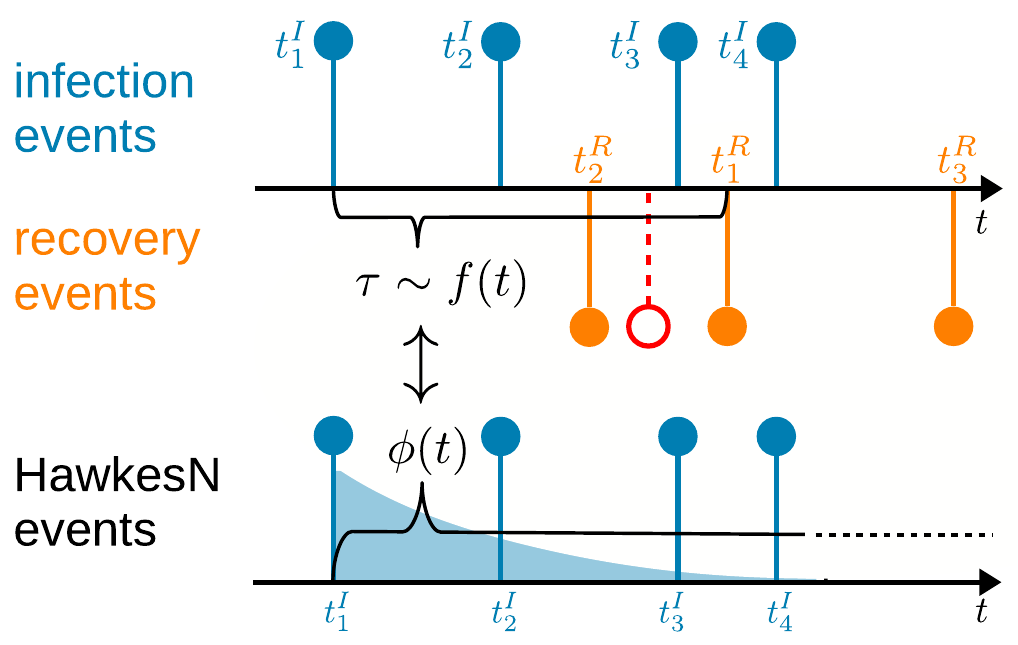}
	\caption{
		A sample stochastic SIR process including an \textcolor{cyan}{infection event history} until time $t$, i.e., $\His^C_t = \{t^I_1, ..., t^I_4\}$, and \textcolor{orange}{recovery events} $\{t^R_2, t^R_1, t^R_3\}$. 
		Infected individuals recover at time intervals $\tau$ following a distribution $f(t)$. 
		The bottom plot presents a corresponding realization of HawkesN events. 
		HawkesN events generate descendants \revA{with the} intensity rate $\phi(t)$. 
		A connection between $f(t)$ and $\phi(t)$ is explored when $f(t)$ is assumed \revA{of} arbitrary parametric forms. 
		The \textcolor{red}{red color} marks an invalid recovery event given upcoming infections.
	}
	\label{fig:invalid-sir-example}
	\vspace{-3mm}
\end{figure}

\section{Preliminaries} \label{sec:preliminaries}

\revA{In this section,} we discuss two classes of stochastic event models, and highlight \revA{the} missing link between them.

\textbf{SIR models}, originally proposed by~\citet{kermack1927contribution}, describe the number of people infected by an epidemic in a fixed population over time.
\revA{The} name \revA{stands for the} three possible states for individuals
--- those in a {\bf S}usceptible state can get {\bf I}nfected, and those infected will eventually {\bf R}ecover or be {\bf R}emoved, and \revA{they are} no longer prone to \revA{the} infection.
The {\em stochastic} variant of \revA{the} SIR model~\citep{bartlett1949some} is concerned with individual state changes,
rather than expected volumes of \revA{individuals} in each state.
The transition of individuals from susceptible to infected is described \revA{by} the {\em infection process}, and that from infected to recovered \revA{by} the {\em recovery process}.

One can  represent \revA{the} stochastic SIR in a fixed population of size $N$ as two sets of random event times, for \revA{the} infections and recoveries, respectively.
Let $\His^C_t$ denote the set of infection event times that happened before time $t$,
and $C_t = |\His^C_t|$ is the number of infection events up to time $t$. Let $i$ index individuals in accordance with their infection time sequence, then $\His^C_t = \{t_i^I \mid  t_1^I = 0, t_1^I<\ldots<t_{C_t}^I<t\}$. The short hand {\em C} stands for {\em cumulative}, i.e., $\His^C_t$ and $C_t$ are not affected by the random events of individuals recovering.
Similarly, let $\His^R_t = \{t_j^R \mid 0<t_j^R<t, t_j^R > t_j^I \}$ denote a set of recovery event times before time $t$, and let $R_t=|\His^R_t|$ be the number of individuals recovered by time $t$. 
We use $U()$ to denote the (index) set of individuals in an event history $\His$.
It follows from the sequential indexing that $U(\His^C_t) = \{1,2,\ldots,C_t\}$, and that the set of recovered individuals is a subset of those infected $U(\His^R_t) \subset U(\His^C_t)$.
We use $\His^I_t$ to express the set of infection event times of infected individuals who have {\em not} recovered by time $t$, i.e., $\His^I_t = \{t^I_j \mid  t^I_j < t, t^R_j > t\}$, and $I_t=|\His^I_t|$. It is easy to see that the still infected set complements the recovered set $U(\His^C_t) = U(\His^R_t) \cup U(\His^I_t)$, 
and $C_t=R_t + I_t$.
\cref{fig:invalid-sir-example} \revA{shows} an example of a stochastic SIR process. 
Based on the definitions \revA{above we have}: $\His^C_t = \{t^I_1, t^I_2, t^I_3, t^I_4\}$, $\His^R_t = \{t^R_2, t^R_1, t^R_3\}$, $\His^I_t = \{t^I_4\}$, $C_t = 4$, $R_t = 3$, $I_t = 1$, \revA{at time $t = t_3^R + \epsilon$. }
\revA{The }{\em susceptible} individuals \revA{are} the ones \revA{who} have \revA{never} been {\em infected}, namely are currently neither {\em infected} nor {\em recovered}: $S_t=~N-C_t=~N-I_t-R_t$.

\revA{The} stochastic SIR \revA{process} is defined by an infection event intensity function $\lambda^I(t)$ and a recovery event intensity function $\lambda^R(t)$~\citep{Yan2008} 
\begin{align}
    \lambda^I (t) = \beta \frac{S_t}{N} I_t ; \,\,\,\, %
    \lambda^R (t) = \gamma I_t \label{eq:lambdaR}
\end{align}
where $\beta$ and $\gamma$ are known as the infection rate and the recovery rate \revA{in SIR terminology}.
The total infection rate is proportional \verifyK{to} the susceptible population $S_t$ and the infected population $I_t$.
Each infected individual recovers independently with the same recovery rate $\gamma$, hence the total recovery rate is proportional to the size of the infected population.
It is also assumed that the recovery process is {\em simple}~\citep{Daley2008}, i.e., only one infection or recovery event can happen in any infinitesimal time interval.

Consider the random variable \textit{recovery time} --- the elapsed time between an individual's infection and recovery. \cref{eq:lambdaR} implies that the \textit{recovery time} is exponentially distributed $f(t) = \gamma e^{-\gamma t}$~\citep{Yan2008}. 

\textbf{Hawkes processes} are \revA{a} type of \revA{self-exciting} point processes, \revA{i.e. processes in which} the occurrence of events increases the likelihood of future events~\citep{hawkes1971spectra}. 
This property is \revA{modeled via the} intensity function:
\begin{equation}
    \lambda(t) = \mu + \sum_{t_i < t} \phi(t - t_i)
\end{equation}
where $\mu$ is the background intensity, \revA{and} $\phi: \Real^+ \rightarrow \Real^+$ is \revA{known as the triggering} kernel \revA{--- the rate of new events generated by} event $t_i$ --- and the summation aggregates \revA{the} influences \revA{of} all past events. 

\textbf{HawkesN process} is a finite-population variant of the Hawkes process~\citep{Rizoiu2017c}. 
Assuming the diffusion \revA{occurs} in a fixed population of size $N$, the event intensity is modulated by the proportion of remaining population:
\begin{align} \label{eq:hawkesn}
    \lambda^H(t) =  \frac{N - N_t}{N} \sum_{t_i < t} \phi(t - t_i) 
\end{align}
$N_t$ is the number of events up to time $t$, the background intensity $\mu$ is \revA{set to} zero and the first event happens at time $0$, i.e., $N_0 = 1$.

\revA{The} stochastic SIR and Hawkes processes have been developed by separate scientific communities for modeling different natural phenomena (epidemics and financial transactions/earthquakes, respectively). 
It is desirable to connect these apparently disparate tools using the common language of stochastic point processes. 

\setlength{\textfloatsep}{5mm}

{\renewcommand{\arraystretch}{1.2}
\begin{table*}[tbp]
	\caption{
		Examples of HawkesN kernel functions $ \phi(t) $, the corresponding SIR recovery time distributions $ f(t) $ and hazard functions $ h(t) $ following 
		Eqs. (\ref{eq:f-to-phi})(\ref{eq:equivalence-f-phi})(\ref{eq:hazard_in_phi}).
		Parameter ranges: $\theta>1$ for Tsallis Q-Exponential kernel, $ \kappa > 0 $, $ \theta > 0 $, $c > 0$ for all others.
	} 
	\label{tab:decay-function-equivalence}
	\centering
	\setlength{\tabcolsep}{8pt}
	\begin{tabular}{lllll}
	  \toprule
	  HawkesN & HawkesN Kernel & SIR Recovery Time & SIR Recovery & Time  \\
	  Kernel Name &  Function $\displaystyle \phi(t)  $  & Distribution $\displaystyle f(t) $ &  Hazard $\displaystyle h(t) $ & Constraint $t$ \\
 	 \midrule
       Linear & $\displaystyle -\kappa\theta t + \kappa $ & $\displaystyle \theta $ & $\displaystyle  \frac{\theta}{-\theta t + 1} $ & $\displaystyle (0, \frac{\kappa}{\theta})$\\ [2ex]%
       Quadratic & $\displaystyle \kappa\frac{\theta^2}{4}t^2 - \kappa\theta t + \kappa$ & $\displaystyle -\frac{\theta^2}{2} t + \theta $ & $\displaystyle \frac{\theta^2 t - 2 \theta}{\theta^2 t^2 - 4\theta t + 4}$ & $\displaystyle (0, \frac{2}{\theta})$ \\ [2ex]
       Gaussian & \( {\kappa e^{-\frac{t^2}{2\theta^2}}}\) & $\displaystyle \frac{t}{\theta^2} e^{-\frac{t^2}{2\theta}} $ & $\displaystyle \frac{1}{\theta^2} t $ & $\displaystyle (0, \infty)$ \\ %
	   Tsallis Q-Exponential~\citep{lima2018hawkes} & $\displaystyle \kappa \bracket{1 + (\theta - 1)t}^{\frac{1}{1 - \theta}} $ &  $\displaystyle  \bracket{1 + (\theta - 1)t}^{\frac{\theta}{1 - \theta}} $  & $\displaystyle  1 + (\theta - 1)t $& $\displaystyle (0, \infty)$ \\ [1ex] %
		Exponential~\citep{hawkes1971spectra} & $\displaystyle \kappa \theta e^{-\theta t} $ & $\displaystyle \theta e^{-\theta t} $ & $\displaystyle \theta $ & $\displaystyle (0, \infty)$ \\ %
        Power-law~\citep{Mishra2016FeaturePrediction} & $\displaystyle \kappa (t + c)^{-(1+\theta)} $ &  $\displaystyle c^{1 + \theta}(1+\theta)(t + c)^{-(2+\theta)} $  & $\displaystyle \frac{1 + \theta}{t + c} $ & $\displaystyle (0, \infty)$ \\ %
 	  \bottomrule
    \end{tabular}
\end{table*}
}

\section{Linking SIR and HawkesN}\label{sec:linking_sir_hawkes}
First, we present a generalized stochastic SIR model with an arbitrary recovery time distribution, and \revA{next} we reveal the connection between the general stochastic SIR and HawkesN. 
Finally, we extend the generalized SIR model with concepts from \revA{the} Hawkes models.

\subsection{SIR with general recovery distributions}
\label{ssec:general_SIR}

\revA{As discussed in \cref{sec:preliminaries}, the} stochastic SIR process implicitly assumes that recovery times \revA{of infected} individuals are exponentially distributed.
\revA{Here we relax this assumption by letting recovery times follow}
an arbitrary distribution $f(t)$.
The recovery intensity for each individual \revA{is given by} the hazard function $h(t)$~\citep{cox1984analysis}, i.e., the recovery time \verifyK{distribution conditioned on recovering after time $t$}:
\begin{equation}
    h(t) = \frac{f(t)}{\int_{t}^{\infty} f(\tau)d\tau}
    \label{eq:hazardfn}
\end{equation}
Considering that individuals recover independently, the overall recovery event intensity is the superposition of recovery intensities of the individuals still infected at time $t$:
\begin{equation} \label{eq:lambdar_general}
    \lambda^R(t) = \sum_{t^I_i \in \His^I_t} h(t - t^I_i) = \sum_{t^I_i \in \His^I_t} \frac{f(t - t^I_i)}{\int_{t-t^I_i}^{\infty} f(\tau)d\tau}
\end{equation}
The overall infection event intensity \revA{remains unchanged} as in \cref{eq:lambdaR}.
Note that, when $f(t)$ is the exponential distribution, \revA{\cref{eq:lambdar_general} simplifies to the infection intensity of the classic SIR in} \cref{eq:lambdaR}.

\revA{Despite being rather straightforward,} to the best of our knowledge, \revA{this is the first work presenting this generalized SIR with arbitrary recovery distributions.}

\subsection{Marginalizing over recovery events}
\label{sec:marginalizing}

One \revA{of the} challenges for using \revA{the} SIR model for social media diffusions is that the definitions of infection and recovery are not straightforward. 
Infection events can be interpreted as posting, sharing or retweeting, \revA{and they} are usually recorded in data traces; 
recovery events can be the times when these posts or discussion topics lose traction, which are rarely \revA{directly} observable. 
This observation implies that one may treat recovery events as {\em latent}, and examine the expected process after marginalizing over them.

We use $\E_{\{t^R_i \mid \His^C_t\}}\bracket{\lambda^I(t)}$ to denote the expected infection intensity over all recovery event times 
up to time $t$:
\begin{align*}
    \E_{\{t^R_i \mid \His^C_t\}} \bracket{\lambda^I(t)}
    &\stackrel{\text{(a)}}{=}  \beta \frac{S_t}{N} \sum_{t^I_{i} \in \His^C_t} \int^\infty_{t - t^I_{i}}f(\tau_i \mid  \His^C_t) d\tau_i \\
    &\stackrel{\text{(b)}}{\geq}  \beta \frac{S_t}{N} \sum_{t^I_{i} \in \His^C_t} \int^\infty_{t - t^I_{i}}f(\tau_i) d\tau_i \numberthis \label{eq:derivation}
\end{align*}

Eq.~(\ref{eq:derivation}a) \revA{follows from ~\citet{Rizoiu2017c}}. 
\revA{Step (b) is because, given $\His^C_t$ an infection history observed up to time $t$,}
the recovery event time of the $i^{th}$ individual $t^R_i$ ($i \in U ( \His^C_t )$) is dependent on the \revA{entire} $\His^C_t$. 
\cref{fig:invalid-sir-example} illustrates this dependence with \revA{the} red \revA{recovery} event being an invalid candidate for $t^R_1$ given \revA{$\His^C_t = \{t^I_1, t^I_2, t^I_3, t^I_4\}$}.
\revA{Intuitively, if the first individual recovers at the time of the red event, there will be zero infected individuals afterwards, rendering impossible the rest of the diffusion.} 
We simplify the dependence using the inequality in Eq.~(\ref{eq:derivation}b) to \revA{the} recovery time distribution $f(t)$. 
We show that
\begin{equation}
    \int^\infty_{t - t^I_{i}}f(\tau_i \mid  \His^C_t) d\tau_i \geq \int^\infty_{t - t^I_{i}}f(\tau_i) d\tau_i
\end{equation}
\revA{with the left and right terms being equal}
when $t^I_i = \max\{\His^C_t\}$. The proof is detailed in \revA{the online supplement}~\citep[appendix A]{appendix}.

Comparing Eq.~(\ref{eq:derivation}b) and \cref{eq:hawkesn}, both $N-N_t$ (for HawkesN) and $S_t = N - C_t$ \revA{(for SIR)} stand for the size of remaining susceptible population --- hence the scaling factors $S_t/N$ and $(N - N_t)/N$ are equivalent. 
Also, both Eq.~(\ref{eq:derivation}b) and \cref{eq:hawkesn} sum over the infected population, and the integral in Eq.~(\ref{eq:derivation}a) is a function of time since infection $t - t^I_i$.
Therefore, marginalizing \revA{the} recovery events reduces the infection intensity of the stochastic SIR to a lower bound \revA{--- the} HawkesN intensity --- \revA{as} long as the following relationship between the HawkeN kernel and recover time distribution holds:
\begin{equation} \label{eq:f-to-phi}
    \phi(t) = \beta \int_t^{\infty} f(\tau)d\tau
\end{equation}
We can express $f(t)$ in terms of $\phi(t)$. $f(t)$ is a probability density function which implies $f(t) \geq 0$ and $\int_0^{\infty} f(\tau)d\tau = 1$, leading to $\phi(0)=\beta$:
\begin{align} \label{eq:equivalence-f-phi}
    f(t) &= -\frac{\phi'(t)}{\phi(0)}
\end{align}
\rev{where we} assume $\lim_{t\rightarrow \infty}f(t) = 0$.
\cref{eq:f-to-phi} and \cref{eq:equivalence-f-phi}
spell out the closed-form relationship between the recovery time distribution $f(t)$ of the stochastic SIR and the kernel function $\phi(t)$ of the HawkesN process. 
From \cref{eq:f-to-phi}, we note that this relationship only holds when $\phi(t)$ is a monotonically decreasing function.
Incorporating \cref{eq:equivalence-f-phi} into \cref{eq:hazardfn}, we can express the recovery hazard function in terms of the HawkesN kernel:
\begin{align} \label{eq:hazard_in_phi}
    h(t) &= -\frac{\phi'(t)}{\phi(t)}
\end{align}
\revA{Given that} $\phi(t)$ \revA{is} monotonically decreasing, $-\phi'(t)$ and $h(t)$ are non-negative.

Table~\ref{tab:decay-function-equivalence} lists six examples \revA{of HawkesN} kernels, with their corresponding recovery time distributions and recovery hazard functions.
The first three rows show the linear, quadratic, and Gaussian kernels,
followed by \revA{the} Tsallis Q-Exponential kernel used in quantum optics and atomic physics~\citep{lima2018hawkes}. 
The last two examples are \revA{the} exponential kernel function and \revA{the} power-law kernel function, widely used for financial data, geophysics, and information diffusion~\citep{hawkes1971spectra,bacry2015hawkes,Mishra2016FeaturePrediction}.

\noindent{\bf Relation to prior work.}
The relationship presented by \citet{Rizoiu2017c} omits the inequality shown in Eq.~(\ref{eq:derivation}b), and \revA{it} is a special case of the result in this work. 
\revA{Their} reasoning is limited to \revA{the} constant recovery hazard functions and \revA{the} exponentially distributed recovery times, with
$f(t) = \gamma e^{-\gamma t}$ and $\phi(t) = \kappa \theta e^{-\theta t}$. 
The main modeling contribution compared to \citep{Rizoiu2017c,Yan2008}
is a new set of analytical relationships among general recovery time distributions, kernel and hazard functions, in 
Eqs. (\ref{eq:f-to-phi})(\ref{eq:equivalence-f-phi})(\ref{eq:hazard_in_phi}).

\subsection{Marked stochastic SIR}
\revA{In real data and apart from event times,}
additional information about individual events \revA{is available}, such as the user profile of a retweet event \revA{or patient characteristics in epidemics.}
Mathematically, the event history $\His^C_m = \{(t^I_1, m_1), ..., (t^I_n, m_n)\}$ is a sequence of pairs of event times and extra event information also known as {\em event marks}. 
To leverage this information, marked variations of Hawkes process models are proposed to incorporate event marks as a scaling factor of kernel functions~\citep{hawkes1971spectra}. 
This idea leads to a marked variation of \revA{the} HawkesN model, \revA{with the intesity function as:}
\begin{equation}
    \lambda^H_{m}(t) = \frac{N - N_t}{N} \sum_{(t^I_i,m_i) \in \His^I_m(t)}m_i^{\rho} \phi(t - t^I_i)
\end{equation}
where $\rho$ controls a warping effect for the mark. 
\revA{Using the generalized connection introduced in \cref{sec:marginalizing},} we are able to obtain a marked stochastic SIR model, whose infection intensity function is
\begin{equation} \label{eq:marked-infection}
    \lambda^I_m(t) = \beta \frac{S_t}{N} \sum_{(t^I_i,m_i) \in \His^I_m(t)}  m_i^{\rho} 
\end{equation}
\verifyK{where, comparing to \cref{eq:lambdaR}, $I_t$ was decomposed to $\sum_{(t^I_i,m_i)\in \His^I_m(t)} m_i^{\rho}$ to account for the individual mark information.}
\revA{The recovery intensity $\lambda_m^R(t)$ is identical to its unmarked counterpart in \cref{eq:lambdar_general}.}

\subsection{Branching factor for SIR}
The basic reproduction number $R_0$ is an important quantity in epidemic models for determining whether an epidemic \revA{is likely to} occur \cite{allen2008mathematical}. 
This quantity conceptually connects to the branching factor $n^*$ from Hawkes processes which is defined as the expected number of events generated by a single infection event~\cite{Rizoiu2017c}, i.e., $n^* = \int^{\infty}_0 \phi(\tau) d\tau$. 
Building upon this observation and \cref{eq:f-to-phi}, we define $R_0$ for stochastic SIR with a general recovery time distribution as
\begin{equation} \label{eq:branching_factor}
    R_0 = n^* = \beta \int^{\infty}_0 \int^{\infty}_{\eta} f(\tau) d\tau d\eta
\end{equation}
\verifyK{Based on~\citep{newman2018networks}, one can also generalize $R_0$ to $\beta \int^{\infty}_0 \tau f(\tau) d\tau$, but we show in \citep[appendix A]{appendix} that this definition is equivalent to~\cref{eq:branching_factor}.}

For marked variations, this quantity is computed by taking expectation over the distribution of event marks. 
\revA{Particularly}, for retweet cascades \revA{where the} event marks \revA{are the count of user followers}, a power law distribution $P(m) = (\alpha - 1) m^{-\alpha}$ \revA{of exponent $\alpha = 2.016$ is determined by \citet{Mishra2016FeaturePrediction}}.
We obtain
\begin{equation}
    R_0 = n^* = \beta \frac{\alpha - 1}{\alpha - 1 - \rho} \int^{\infty}_0 \int^{\infty}_{\eta} f(\tau) d\tau d\eta
\end{equation}
We refer to this quantity as just the branching factor $n^*$ in the following sections to avoid confusion.

\begin{algorithm}[tb]
\caption{Simulating generalized stochastic SIR}
\label{alg:simulation}
\begin{flushleft}
\textbf{Input}: Recovery time distribution $f(t)$, parameters $\{N, \beta\}$
\textbf{Output}: Infection event times $\His^C$ and recovery event times $\His^R$
\end{flushleft}
\begin{algorithmic}[1] %
\STATE Set current time $T= 0$.
\STATE Initialize $\His^C = \{0\}$ with one initial infection at time $0$.
\STATE Initialize $\His^R = \{\eta \}$ where $\eta \sim f(t)$ and $t^R_1 = \eta$.
\WHILE{$|\His^C| < N$}
\STATE $s = - \frac{log(u)}{\lambda*} $ where $u \sim U(0, 1)$
\STATE Compute $\Lambda^I(t) = \int^t_0 \lambda^I(\eta) d\eta$ from $\His^C, \His^R$
\STATE $T = T + (\Lambda^I)^{-1}(s)$
\IF {$T = \infty$}
\STATE \textbf{break} \textit{// No infection will occur}
\ELSE
\STATE $\eta \sim f(t)$ \hspace{1.5mm} \textit{// Draw recovery time, update histories}
\STATE $\His^R = \His^R \cup \{T + \eta \}, \His^C = \His^C \cup \{T\}$
\ENDIF
\ENDWHILE
\STATE \textbf{return} $\His^C,\His^R$
\end{algorithmic}
\end{algorithm}

\section{A Set of Tools for Stochastic SIR}
\label{sec:simulation_estimation}
In this section, we introduce a set of tools for the stochastic SIR with general recovery time distributions and HawkesN, enabling one to simulate event realizations, estimate model parameters, assess fitted results and predict final diffusion sizes.

\noindent{\bf Generalized SIR simulation.} 
\revA{The generalized SIR proposed in \cref{eq:lambdar_general} cannot be simulated using the approach described by ~\citet{allen2008mathematical}}
as the recovery event rate is no longer piece-wise constant. 
We show a procedure of sampling general stochastic SIR processes, by sampling each infection event and its corresponding recovery time.

Starting from the first infection event at $t=0$, \cref{alg:simulation} iterates between two steps. 
Step one is to sample the recovery event time according to $f(t)$ (line 11-12), \revA{step two} is to sample the next infection time by the random time change theorem~\citep{laub2015hawkes} (line 5-7). 
Specifically, \verifyK{because future recovery times have been sampled for existing infection events}, the infection event intensity can be then derived from \cref{eq:lambdaR} as a piece-wise constant function. 
The infection intensity leads to analytical forms of the cumulative infection intensity $\Lambda^I(t) = \int^t_0 \lambda^I(s) ds$ and its inverse $(\Lambda^I)^{-1}(\cdot)$. 
It is presented that $(\Lambda^I)^{-1}(\cdot)$ can convert a time interval sampled from a Poisson process with unit rate (line 5) to an interval generated by the intensity function $\lambda^I(t)$ (line 7)~\citep{laub2015hawkes}.
The process terminates when all $N$ individuals have been infected (line 4), or when the infection rate falls to zero (line 8).

\noindent{\bf Parameter estimation.} 
We use maximum likelihood to estimate model parameters given event history via standard optimization packages.
The likelihood functions of stochastic SIR and HawkesN can be derived 
from the general likelihoods for point processes~\citep{Daley2008}.
Details are in \revA{the online supplement}~\citep[appendix B]{appendix}.

Suppose events are generated with an underlying stochastic SIR model.
To estimate its parameters,
when both infection and recovery events are observed, 
the stochastic SIR likelihood is maximized;
when only infection events are observed, we estimate with the HawkesN likelihood \verifyK{to account for their latent recovery information}.
Due to the inequality in \cref{eq:derivation}, the HawkesN likelihood is a biased estimator for stochastic SIR \revA{process parameters}. 
\revA{We study this bias in \cref{subsec:synthetic_data} and}
we show empirically that \revA{it} reduces as \revA{the branching factor increases}.

\noindent{\bf Goodness-of-fit assessment.}
\revA{Given that the generalized SIR model can accommodate a wide range of recovery distribution functions, one natural question is how to assess the fitness of}
fitted models \revA{to} observed events, \revA{choose between different parametric families} and provide a guide to \revA{predict} future events \citep{chen2018marked}. 
Due to the aforementioned random time change theorem~\citep{laub2015hawkes}, for observed infection events $t^I_i \in \His^C_t$ correctly described by an infection intensity function $\lambda^I(t)$, the cumulative infection intensities between infection events are time intervals generated from a Poisson process with unit rate or, equivalently, follow a unit rate exponential distribution:
\begin{equation} \label{eq:goodness-of-fit}
    \mathcal{T}_i = \int^{t^I_i}_{t^I_{i-1}} \lambda^I(\tau) d\tau,\hspace{0.5cm} \mathcal{T}_i \sim e^{-t}
\end{equation}
Three statistical tests are applied:
\revA{the} Kolmogorov-Smirnov (KS) test and
\revA{the} Excess Dispersion (ED) test \revA{to} measure the significance of the proposition $\{\mathcal{T}_i\} \sim e^{-t}$; 
\revA{the} Ljung-Box (LB) test \revA{to} determine the independence among $\{\mathcal{T}_i\}$. 

\citet{lallouache2016limits} note that \revA{the} KS test is a more demanding test than \revA{the} ED test. 
Specifically, \revA{the} KS test evaluates the empirical cumulative density function (CDF) of $\{\mathcal{T}_i\}$ against the theoretical CDF of the unit rate exponential distribution (i.e., $1 - e^{-t}$) producing two values: a p-value, indicating the significant level of $\{\mathcal{T}_i\}$ not being drawn from the nominated theoretical CDF, and a distance $D$ between the empirical CDF and the theoretical CDF~\cite{massey1951kolmogorov}. 
As models presented in this paper are evaluated against the same theoretical CDF, we employ this distance measure $D$ as a fitting performance metric for model comparison.

\begin{figure*}[!htp]
    \centering
    \begin{subfigure}{0.496\textwidth}
    \includegraphics[width=\textwidth]{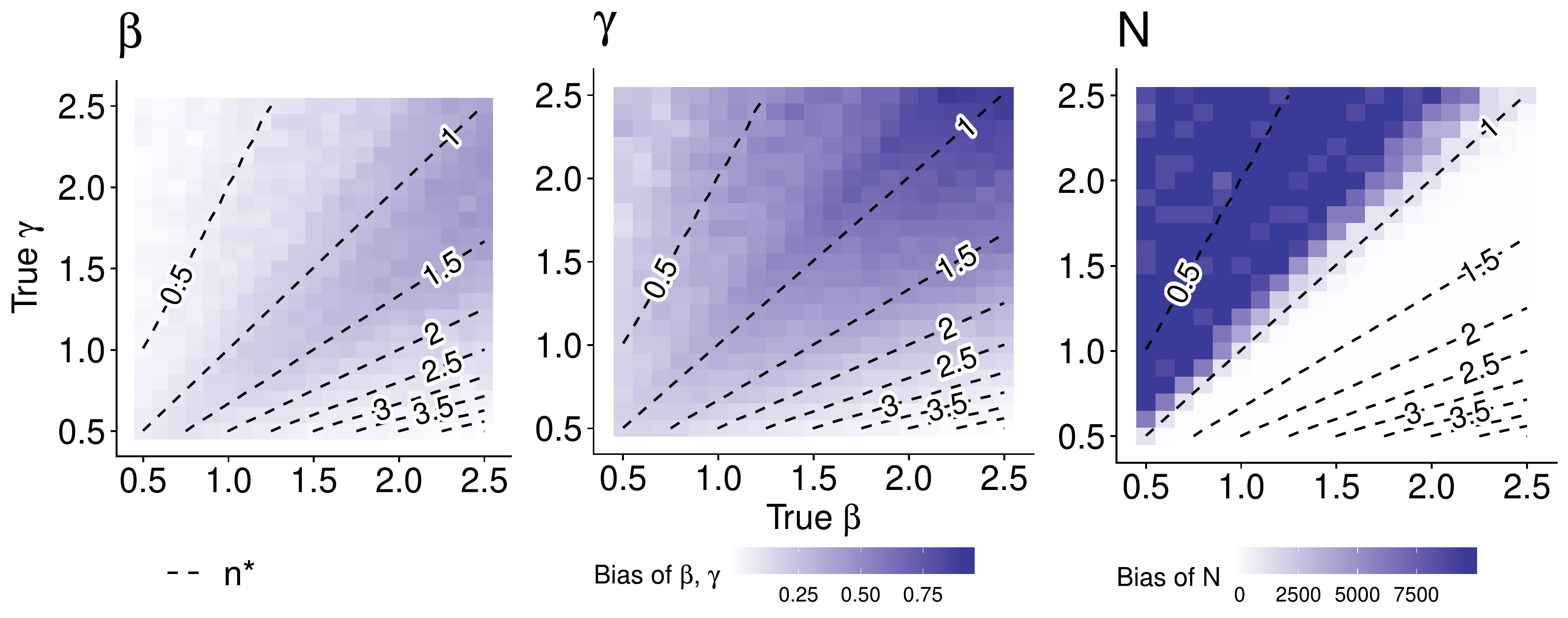}
    \vspace{-4mm}
    \caption{Exponential}
    \vspace{-4mm}
    \label{subfig:parameter-EXPN}
    \end{subfigure}
    \hspace*{\fill}
    \begin{subfigure}{0.496\textwidth}
    \includegraphics[width=\textwidth]{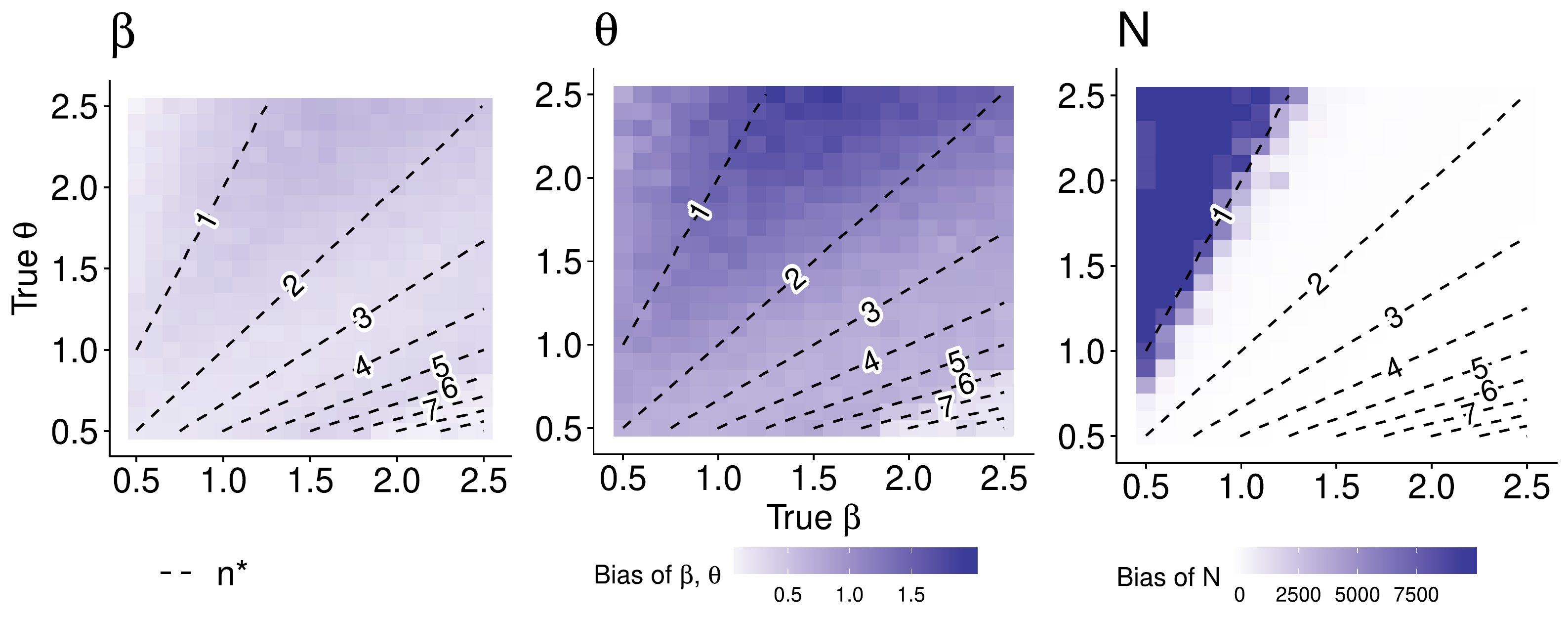}
    \vspace{-4mm}
    \caption{Power-law}
    \vspace{-4mm}
    \label{subfig:parameter-PLN}
    \end{subfigure}
    \caption{
        Bias of estimating parameters with HawkesN likelihood functions on simulated stochastic SIR infection events. Stochastic SIR with an exponential \textbf{(a)} and a power-law \textbf{(b)} recovery time distributions are evaluated. Chosen parameters are: \textbf{(a)} $N = 200, \beta = \gamma = \{0.5, 0.6, ..., 2.5\}$; \textbf{(b)} $N = 200, c = 2, \beta = \theta = \{0.5, 0.6, ..., 2.5\}$. Estimation bias is computed with \textit{absolute errors} --- lighter colors indicate lower bias. The dotted contour lines are the branching factors given the parameter sets.
    }
	\label{fig:parameter}
\end{figure*}

\noindent{\bf Diffusion final size prediction}
Point processes are generally applied for event history explanation and not optimized for prediction. 
To predict the diffusion final size (a.k.a the popularity for a Twitter cascade), we follow~\citep{Mishra2016FeaturePrediction} \verifyK{by using a regression layer on top of the proposed models}.
\revA{We predict} a quantity $\sigma$ which can be interpreted as the proportion of remaining population that will be involved in the diffusion, i.e.,
\begin{equation}
    \hat{C}_{\infty} = C_t + \sigma (N - C_t)
\end{equation}
where $t$ is the observation time, $C_t$ is the number of cumulative infection events, $N$ is the fitted population size and $\hat{C}_{\infty}$ is the predicted diffusion final size. 
We note that $\sigma > 1$ is possible due to the underestimation of $N$ given observed events or the growth of population as diffusion unfolds. 
We use the fitted parameters and \revA{the} derived branching factor (e.g., $\{\beta, \gamma, \rho, N, n^*\}$ for exponentially recovered stochastic SIR) as features to train a $sigma$ predictor. 
This setup can also be applied to HawkesN given $N_t$ and its fitted parameters. 
The prediction experiment is set up to reproduce experiments in \citep{Mishra2016FeaturePrediction}, and \revA{we further detail it} in \cref{sec:experiments}.

\section{Experiments}\label{sec:experiments}
\revA{We first study the fitting of SIR parameters when the recovery times are not observed, and} 
we design an empirical validation of the connection between the stochastic SIR and the HawkesN models through simulation and parameter estimation (in \cref{subsec:synthetic_data}). 
Next, we investigate the performance of HawkesN models on three large Twitter cascade datasets in terms of goodness-of-fit, holdout log-likelihood and final diffusion size prediction (in \cref{subsec:modelling-twitter-diffusions}).

\noindent{\bf Models and fitting.}
We use the following abbreviations when presenting our results:
\emph{EXP}, \emph{PL} and \emph{QEXP}, stand for Hawkes models with \revA{the} exponential~\citep{hawkes1971spectra}, power-law~\citep{Mishra2016FeaturePrediction} and Tsallis Q-Exponential kernel functions, respectively; 
\emph{EXPN}, \emph{PLN} and \emph{QEXPN}, refer to HawkesN models with corresponding kernel functions;
\emph{SI} is the stochastic \textbf{S}uscepitable-\textbf{I}nfected model as an epidemic model benchmark for comparison.
The estimation of Hawkes models is \revA{performed as} described by \citet{Mishra2016FeaturePrediction}, \revA{i.e., the} model parameters are fitted on \revA{an initial} training part of a cascade through maximizing \revA{the} log-likelihood functions.
\revA{The} log-likelihood functions can be found in \revA{the online supplement}~\citep[appendix B]{appendix} for HawkesN, and in~\citep{Mishra2016FeaturePrediction} for Hawkes.
The parameter learning and simulation of the stochastic SI model can be adopted from the stochastic SIR model with $\gamma = 0$.

\subsection{Fitting SIR parameters with latent recoveries} 
\label{subsec:synthetic_data}

\revA{In many applications, including in epidemiology, the recovery events are unobserved.
It is therefore desirable to be able to fit the SIR model using infections events only.
In this section we show how to achieve this, and}
we \revA{empirically} validate the connection shown in \cref{{sec:marginalizing}} by 
simulating stochastic SIR and retrieving SIR parameters with \revA{the} HawkesN log-likelihood functions with corresponding kernel functions. 
We \revA{construct} a rich set of parameters for stochastic SIR with \revA{the} exponential (Fig.~\ref{fig:parameter}a) and power-law (Fig.~\ref{fig:parameter}b) recovery time distributions.
For each parameter set shown in~\cref{fig:parameter} (each grid cell), we simulate $1000$ stochastic SIR realizations (using \cref{alg:simulation}). 
We hide the recovery events $\His^R_t$ of these realizations and \revA{we} fit HawkesN processes on infection event times $\His^C_t$. 
We jointly fit $100$ realizations \revA{at a time} by \revA{summing their log-}likelihood functions.

\revA{In each} grid cell in \cref{fig:parameter}, \revA{the colors shows} the fitting bias \revA{--- i.e., the} \textit{absolute error} between simulation parameters and the median of fitted parameters.
Note that, for ease of comparison, we have transformed the fitted HawkesN parameters into SIR parameters (using \cref{eq:f-to-phi,eq:equivalence-f-phi}, and \cref{tab:decay-function-equivalence}).
Also we notice in experiments that the power-law kernel as defined in \citep{Mishra2016FeaturePrediction} is over-determined, and we fix $c=2$ both in simulation and in fitting.

Visibly, the bias of $\beta$ is relatively small due to its direct presence in the infection intensity function $\lambda^I(t)$. 
For \revA{the} other parameters, their bias starts \revA{relatively} high \revA{for low values of the} branching factor (upper-left corners in \cref{fig:parameter}, shown as contour lines) and gradually diminishes as the branching factor grows. 
When the branching factor is large (bottom-right corners in \cref{fig:parameter}), the fitted parameters match closely with the simulation parameters.
Processes with large branching factors are commonly of interest (e.g., $R_0 = 18$ for measles in epidemiology~\citep{brauer2008mathematical}). 
For this reason, this evaluation supports the application of HawkesN log-likelihood functions to retrieve SIR parameters when recovery event times are missing and high branching factors are observed.

\subsection{Modeling diffusions on Twitter}
\label{subsec:modelling-twitter-diffusions}

\begin{table}[tbp]
	\centering
	\setlength{\tabcolsep}{4pt}
	\caption{Statistics of the three social media datasets.}
	\vspace{-0.2cm}
	\begin{tabular}{rrrrrr}
	  \toprule
	 & \#cascades & \#tweets & Min. & Mean & Median \\ 
	  \midrule
        \textit{ActiveRT} &  39,970 & 7,873,733 & 20 & 197 & 41 \\ 
        \textit{Seismic} & 166,076 & 34,784,488 & 50 & 209 & 111 \\ 
        \textit{NEWS} & 20,093 & 3,252,549 & 50 & 162 & 90 \\ 
	  \bottomrule
	\end{tabular}
	\label{tab:dataset-profiling}
\end{table}

\begin{figure*}[!tbp]
    \begin{flushleft}
        \begin{subfigure}{0.43\textwidth}
            \includegraphics[width=\textwidth]{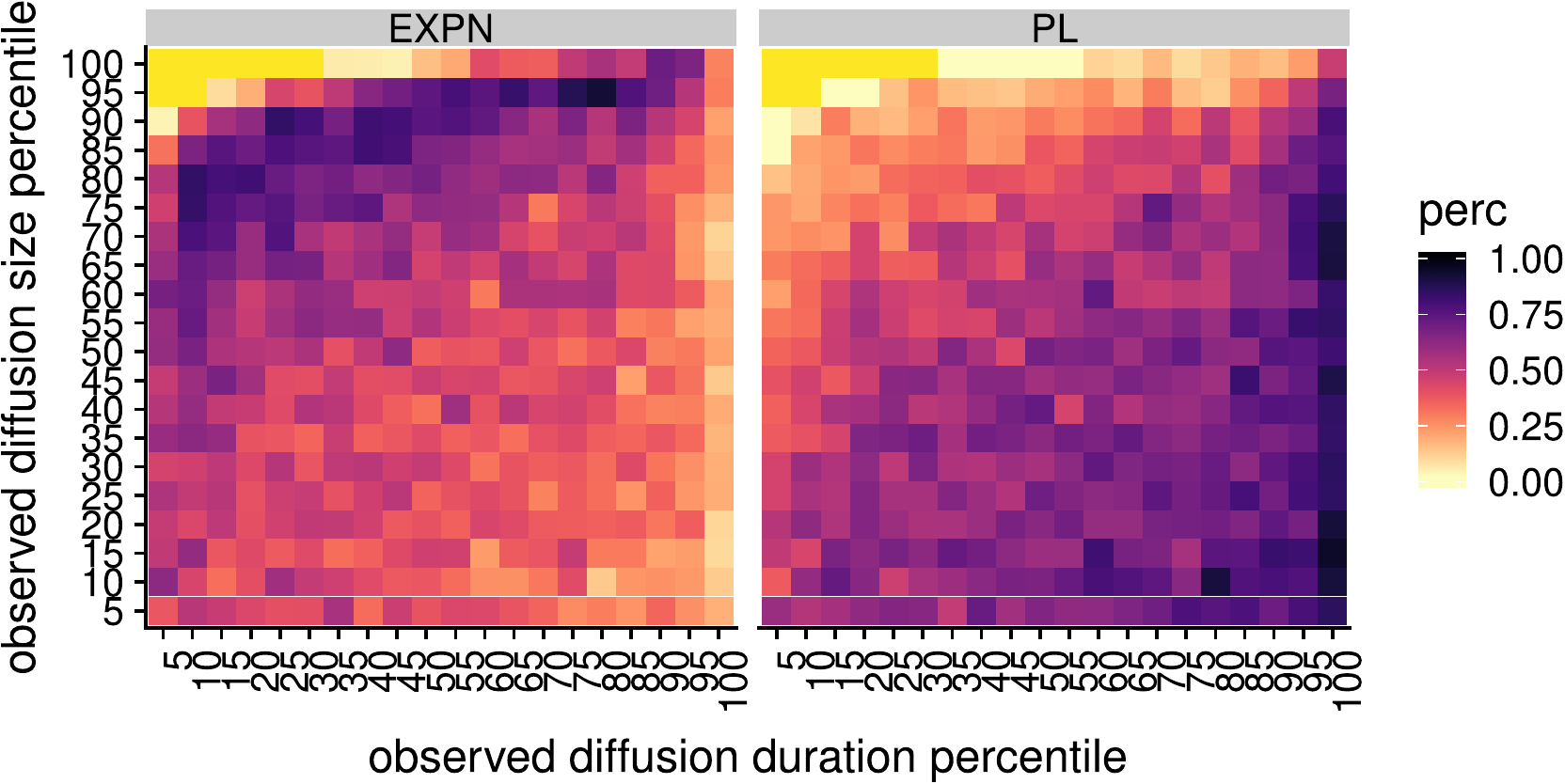}
            \caption{EXPN vs. PL on \textit{NEWS}}
            \label{subfig:news-best-fit-model}
        \end{subfigure}
        \hspace{5mm}
        \begin{subfigure}{0.53\textwidth}
            \includegraphics[width=\textwidth]{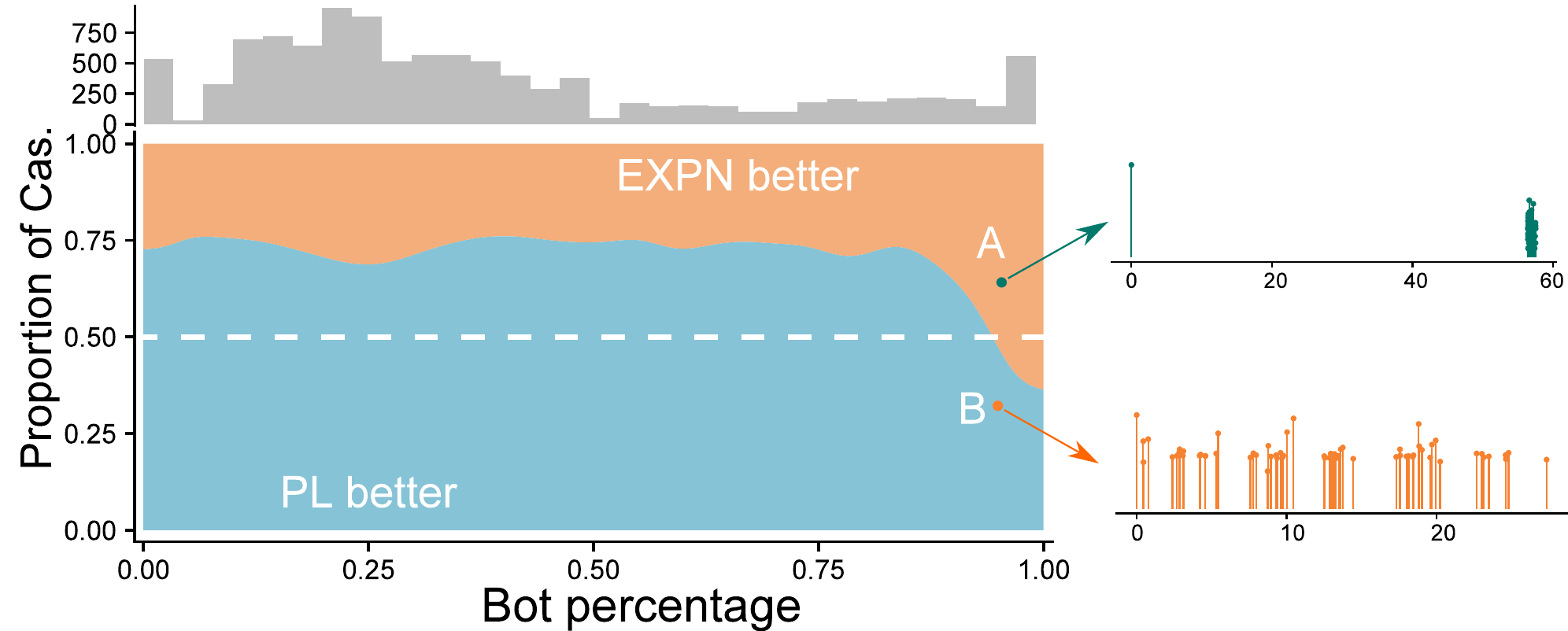}
            \caption{EXPN vs. PL on \textit{ActiveRT} w.r.t. percentage of bots in cascades}
            \label{subfig:bot}
        \end{subfigure}
    \end{flushleft}
    \vspace{-2mm}
    \caption{
        Comparing model goodness-of-fit using KS test values. \emph{(a)} The most distinct model pair, EXPN and PL, on \textit{NEWS} where colors of bins represent proportions of cascades that are better fitted by the model. \textcolor{Goldenrod}{Yellow} means there is no cascade in the bins. \emph{(b)} EXPN compares to PL on \textit{ActiveRT} in terms of the percentage of bots involved in observed retweet cascades. The upper-panel \textcolor{gray}{histogram} counts the number of cascades at different bot percentages; the lower-panel plot depicts proportions of cascades better fitted by \textcolor{orange}{EXPN} or \textcolor{cyan}{PL} at given bot percentages. Two high bot-percetage cascade examples (cascade~\textit{A} and cascade~\textit{B}) better fitted by EXPN and PL, respectively, are shown on the right-hand side.
    }
    \label{fig:botness}
    \vspace{-2mm}
\end{figure*}

\noindent{\bf Datasets.}
We use three publicly available Twitter datasets containing retweet cascade --- individual sequences of retweet events following a single initial tweet.
Each tweet in the cascade is considered to be an infection event, i.e., a cascade is the collection
$\His^C = \{(t^I_1, m_1), (t^I_2, m_2), ...\}$ where $t^I_i \in \His^C$ is the time stamp of the $i^{th}$ retweet in the cascade and $m_i$ is its associated mark information, namely the number of followers of the user.
The \textit{Seismic} dataset was constructed by \citet{Zhao2015SEISMIC:Popularity}, and it contains a subset of all tweets in a month.
The \textit{NEWS} dataset was collected by \citet{Mishra2016FeaturePrediction} by crawling all tweets that contain links to popular news sites, such as New York Times and CNN, for four months in 2015.
The \textit{ActiveRT} dataset\footnote{The total number of cascades in \textit{ActiveRT} is 39,970 rather than 41,411 reported in \citep{Rizoiu2017c} after we filtered out 1,441 duplicate cascades.} was collected by \citet{Rizoiu2016ExpectingPopularity} over 6 months in 2014, by capturing all tweets containing links to Youtube videos.
\cref{tab:dataset-profiling} \revA{summarizes the} three datasets.

\begin{table}[tbp]
	\caption{Goodness-of-fit assessments on three datasets. 
	Models are fitted on initial $40\%$ of each cascade event history with marks. 
	\revA{The numbers in each cell} indicate \revA{the} percentages of \revA{cascades for which each} model \revA{passes the} nominated statistical tests (in \cref{sec:simulation_estimation}) at the $0.01$ significance level. 
	Darker \revA{colors signify a larger} fraction of cascades passing.}
	\vspace{-0.1cm}
	\centering
	\setlength{\tabcolsep}{1pt}
	\renewcommand{\arraystretch}{1.01}
	\begin{tabular}{L{1.2cm}cC{0.9cm}C{0.9cm}C{0.85cm}C{0.8cm}C{0.8cm}C{1cm}C{0.5cm}}
		\toprule
		& Test& EXP & EXPN & PL & PLN & QEXP & QEXPN & SI\\
		\midrule
		\multirow{3}{*}{\textit{ActiveRT}}& KS & \multicolumn{7}{l}{\multirow{9}{*}{ \includegraphics[width=0.355\textwidth]{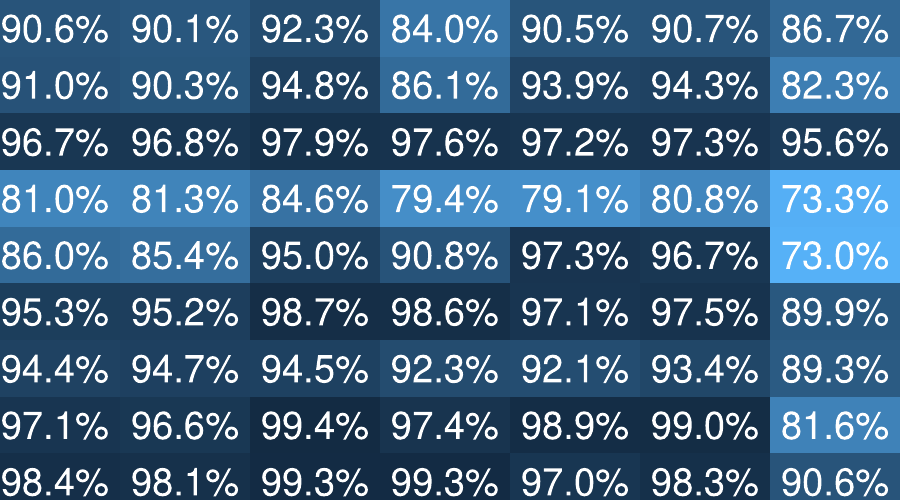}}}\\
		& ED \\
		& LB \\

		\multirow{3}{*}{\textit{Seismic}}& KS \\
		& ED \\
		& LB \\

		\multirow{3}{*}{\textit{NEWS}} & KS \\
		& ED \\
		& LB \\
		\bottomrule
	\end{tabular}
	\label{tab:goodness-of-fit}
\end{table}

\noindent{\bf Goodness-of-fit tests.} 
We first conduct \revA{the} goodness-of-fit tests described in \cref{sec:simulation_estimation} on all three datasets. 
The first $40\%$ of event history of each cascade is used for model fitting.
\cref{tab:goodness-of-fit} shows the percentages of cascades for which each model passes the tests at a $0.01$ significance level. We note that the passing rates of the SI model is consistantly worse than other models due to the model simplicity, so we focus our following experiments and discussions only on other models.

\revA{First, we see that} 
the statistical test on the independence of transformed event times (LB test on $\{\mathcal{T}_i\}$ in \cref{eq:goodness-of-fit}) presents high passing percentages ($97.57\% \pm 1.16\%$) across all models (except SI) and datasets. 
\revA{The other} two tests (KS test and ED test) 
mostly agree on \revA{the performance of models with respect to each other,}
despite KS being a more demanding test.

When comparing 
Hawkes and HawkesN, \revA{we observe} an increase in performance \revA{for the} Tsallis Q-Exponential kernel (from QEXP to QEXPN), and a decrease from PL to PLN. 
EXP and EXPN, on the other hand, share similar performance. 
This indicates \revA{that} the effect of \revA{modulating the} Hawkes intensity \revA{by} a finite population \revA{for modeling} retweet cascades \revA{is dependent on the choice of} kernel.

\noindent{\bf Model goodness-of-fit comparison.} 
By leveraging distances produced in the KS tests, we explore \revA{the} modeling performance differences \revA{for every given} dataset. 
Given two models $M_1$ and $M_2$ that pass KS test on a cascade $\His^C_t$, we assume $M_1$ fits $\His^C_t$ better if it has a lower KS test distance than $M_2$, denoted $D_{M_1}(\His^C_t) < D_{M_2}(\His^C_t)$. 
\revA{Next, we tabulate the cascades in each dataset against two dimensions: cascade duration (the time of the last event) and cascade size (number of events), both in percentiles.}
\revA{\cref{subfig:news-best-fit-model} compares the two models with the higher KS passing rate (EXPN and PL), on the \textit{NEWS} dataset} (refer to~\citep[appendix C]{appendix} for other model pairs and datasets).
Grid cells depict \revA{the} proportions of cascades that are better fitted by \revA{one} model \revA{or the other}.
\revA{Visibly,} EXPN \revA{fits better} cascades with larger diffusion sizes and shorter diffusion durations, whereas PL performs better on less popular cascades with longer durations. 
\revA{This indicates that} PL and EXPN are two complementary models on \textit{NEWS}, \revA{capturing different diffusion dynamics.}
\begin{figure*}[!tpb]
    \centering
    \begin{subfigure}{0.295\textwidth}
        \includegraphics[width=\textwidth,page=2]{images/{2017-07-07-holdout-nll-perc-news-0.40-unmarked}.pdf}
        \caption{Holdout log-likelihood}
        \label{subfig:generalization-news}
        \vspace{-2mm}
    \end{subfigure}
    \begin{subfigure}{0.295\textwidth}
        \includegraphics[width=\textwidth,page=2]{images/{2017-07-07-holdout-nll-perc-news-0.40}.pdf}
        \caption{Holdout log-likelihood with marks}
        \label{subfig:generalization-activert}
        \vspace{-2mm}
    \end{subfigure}
    \begin{subfigure}{0.362\textwidth}
        \includegraphics[width=\textwidth]{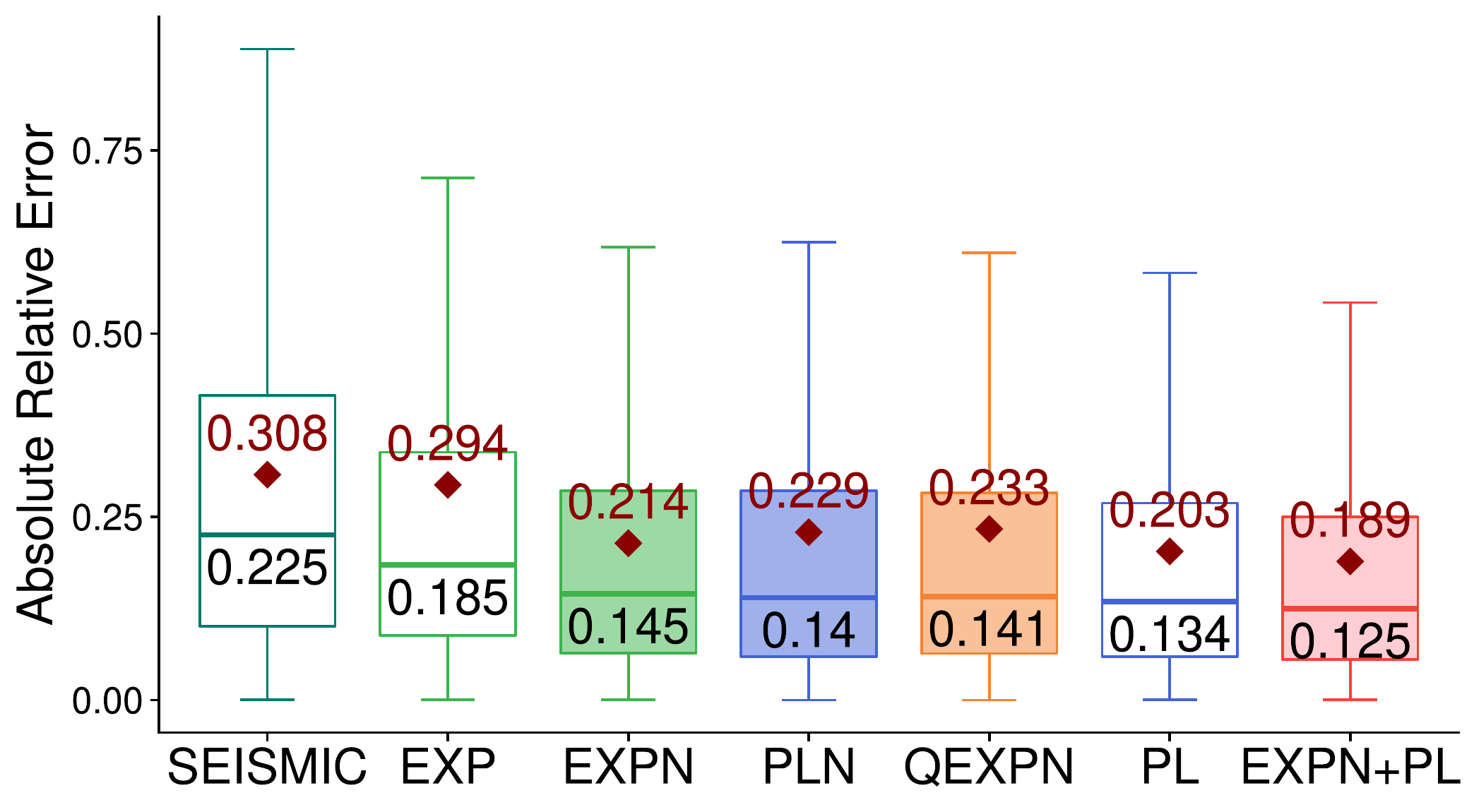}
        \caption{Popularity prediction performance}
        \label{subfig:prediction-news}
        \vspace{-2mm}
    \end{subfigure}

    \caption{
        Fig.~\textit{(a)-(b)} depict holdout negative log-likelihood per event
        of six models on \textit{NEWS}, with and without additional mark information.
        Fig.~\textit{(c)} shows diffusion final popularity prediction performance on \textit{NEWS}.
        The red diamond shows the mean value in each boxplot --- lower is better.
    }
    \vspace{-3mm}
	\label{fig:holdout-experiments}
\end{figure*}

\noindent{\bf Linking modeling to botness.} 
\revA{Here, we investigate a possible factor that induces the retweet dynamics that are better captured by EXPN compared to PL: non-human participation in cascades.}
\verifyK{We choose to analyze \textit{ActiveRT} where the user information of individual events is available.}
We use the Botometer API~\citep{yang2019arming} to identify Twitter bots and we collect data for $1,174,248$ unique users involved in the first $40\%$ event history of the $39,549$ cascades in \textit{ActiveRT}. 
Due to the rate limit of the API, we only crawled cascades that have less than $2,500$ events. 
Given a user $i$, there are three possible outcomes from the API: 
a botness score $b_i \in [0, 1]$ of the user; 
when the user has a private profile; 
or when the user was suspended by Twitter. 
As this data was collected $5$ years after the creation of \textit{ActiveRT} in 2019, we assume users suspended by Twitter are bots. 
Eventually, we classify users who have been suspended or who have $b_i \geq 0.6$ as bots~\citep{rizoiu2018debatenight}.

\revA{First, we group the cascades based on the proportion of bots that participate in each of them (in percentiles).
For each percentile bin,}
\cref{subfig:bot} displays the proportions of cascades better fitted by EXPN and PL.
We only keep cascades that satisfy $|D_{EXPN}(\cdot) - D_{PL}(\cdot)| \geq 0.05$ to identify \revA{the} cascades significantly better fitted by \revA{each} model. 
This condition filters out $72.96\%$ of cascades suggesting that EXPN and PL show similar performance on most cascades. 
We find 
that, \revA{while for most of the remaining cascades PL fits better than EXPN}, when more than $90\%$ of bots involve in a cascade, \eat{its temporal dynamic is more likely to be explained by EXPN compared to PL}\verifyK{around $60\%$ of the cascades are better fitted by EXPN}.
\revA{In \cref{subfig:bot}, we denote A, the bot dominated cascades better fitted by EXPN, and B, those better fitted by PL, and we show one typical example from each.}
\revA{Cascades in A exhibit densely clustered events, with large intervals of no activity between them, whereas the cascades in B have the events more evenly spread.}
Intuitively, the temporal behavior of cascades \revA{in} A tends to be more bot-driven, \revA{as bots retweet each other in rapid sequences and with small delays}.

\noindent{\bf Generalization to unseen data.}
On each of the three dataset, we fit the parameters of all six models. 
We follow the experimental setup in \citep{Rizoiu2017c}:
$40\%$ of the tweets in each cascade are used to fit model parameters, and we report the negative log-likelihood on the remaining $60\%$ of the events normalized by the event count.

Fig.~\ref{fig:holdout-experiments}a and b show as boxplots the generalization performance on \textit{NEWS}, \revA{without and with marks respectively}.
Two conclusions emerge.
First, the power-law kernel for both Hawkes and HawkesN consistently outperforms \revA{other kernel functions}.
This emphasizes the importance of developing \revA{the generalized SIR} model, \revA{as different types of parametric kernel function might fit different types of data better}.
Second, HawkesN outperforms Hawkes confirming results reported in \citep{Rizoiu2017c}.
\revA{The results on the other datasets depict a very similar conclusion, and they are shown in the online supplement}~\citep[appendix C]{appendix}.

\noindent{\bf Popularity prediction.} 
\revA{We predict final retweet cascade popularity following the setup described in}
~\citep{Mishra2016FeaturePrediction}.
We \revA{observe each cascade for one} hour \revA{and we} fit model parameters; \revA{we} predict final diffusion sizes (popularity) \revA{and test against the observed final cascade size.
We measure performance using the} Absolute Relative Error (ARE):
\begin{equation*}
    ARE = \frac{|\hat{N}_{\infty} - N_{\infty}|}{N_{\infty}}
\end{equation*}
where $\hat{N}_{\infty}$ and $N_{\infty}$ are the predicted size and the true size, respectively.
We compare HawkesN models to \revA{the Hawkes models (}EXP, PL), and to SEISMIC~\citep{Zhao2015SEISMIC:Popularity}. 
We use the GBM package in R~\citep{gbm} to train the $\sigma$ predictor described in~\cref{sec:simulation_estimation}. 
Furthermore, we adopt the observation that EXPN and PL are two complementary models on \textit{NEWS} to introduce a combined model by averaging EXPN and PL prediction outcomes~\citep{zhou2012ensemble}.
Results are reported with 10-fold cross validation where 6 folds are used for testing after trained on 4 folds during each iteration.

\revA{\cref{subfig:prediction-news} shows the prediction performances on the \textit{NEWS}.
The performances on \textit{Seismic} (also employed by \citet{Mishra2016FeaturePrediction}, where it is called ~\textit{TWEET-1MO}) and on \textit{ActiveRT} are shown in the online supplement~\citep[appendix C]{appendix}.}
Among \revA{all the} Hawkes and HawkesN models, PL delivers the best prediction performance, and EXPN predicts better than EXP. 
These observations align \revA{with analyses in the} previous \revA{sections}. 
Overall, the combined model, EXPN+PL, \revA{consistently} outperforms all other models, \revA{on all datasets.} 
It provides a choice to deal with complementary modeling power of kernel functions on different cascades. 
\revA{This only reinforces the conclusion that there may exist more than one cascade dynamics, and that each model captures the best one of them.}

\section{Discussion and conclusion}
In this work, we \revA{introduce} a connection between generalized stochastic SIR models and self-exciting point processes in a finite population.
The connection stems from the relationship between the recovery time distributions in SIR and \revA{the} kernel functions in HawkesN processes. 
In addition, we develope \revA{algorithms for}
simulation, parameter estimation and evaluation \revA{for the} stochastic SIR processes and the corresponding HawkesN processes. 
The modeling insights and the computational tools describe a rich set of self-exciting kernel functions,
and they are more general than traditional stochastic SIR with piece-wise constant rates. 
In fact, it describes SIR with arbitrary recovery time distributions, 
and monotonically decreasing Hawkes kernels.
We compare models with three kernel functions --- an exponential, a power-law and a Tsallis Q-Exponential --- on three large Twitter retweet cascade datasets. %
We observe differences in model performance in terms of goodness-of-fit tests. 
Final popularity prediction is improved by combining two complementary models.

\noindent{\bf Limitations and future work} 
Non-monotonically decreasing kernel functions, such as the Rayleigh function, have been used in the point process literature~\citep{ding2015video,Mishra2016FeaturePrediction,rodriguez2011uncovering}. 
Although it cannot be linked to the CCDF of recovery events in epidemics, the intuition of the Rayleigh function \revA{stems from} the concept of the disease incubation period in epidemiology. 
We plan to broaden the connection, e.g., between HawkesN and variants in the epidemic models family.

In general, this newly established bridge between distinct classes of stochastic point processes opens up many research topics such as using modern machine learning tools to design objectives and estimation procedures, 
causal inference in epidemic models, 
and novel applications of either model in new data domains.

\section*{Software and Runtime Information}
The simulation and fitting algorithms described in~\cref{sec:linking_sir_hawkes} have been implemented as an R package available on \url{http://bit.ly/34qiDTK} and the code for reproducing experiments in~\cref{sec:experiments} can be found on \url{https://bit.ly/3697ojK}. Fitting HawkesN models on a $500$ event cascade takes $4.8$ minutes in average with $10$ parallel random parameter initializations on $2.2$GHz cpus where the likelihood function evaluation complexity is quadratic in the number of events.

\subsection*{Acknowledgments}
\small{This research is supported by Asian Office of Aerospace Research and Development (AOARD) Grant 19IOA08, Australian Research Council Discovery Project DP180101985 and the Data61, CSIRO PhD scholarship. We thank our reviewers, Swapnil Mishra and other lab members at ANU CMLab for their helpful comments. We also thank the National Computational Infrastructure (NCI) for providing computational resources, supported by the Australian Government.}
\bibliographystyle{ACM-Reference-Format}
\bibliography{acm}

\clearpage
\appendix
\onecolumn

Accompanying the submission \textit{\titlename}.
\section{Linking SIR to Hawkes}
\subsection{Detailed derivation of the inequality between generalized stochastic SIR and HawkesN}
We use $\E_{\{t^R_i | \His^C_t\}}\bracket{\lambda^I(t)}$ to denote the expected infection intensity over all recovery event times of infected individuals up to time $t$:
\begin{align*}
    & \E_{\{t^R_i | \His^C_t\}} \bracket{\lambda^I(t)} \\
    &\stackrel{\text{(a)}}{=} \beta \frac{S_t}{N} \E_{\{t^R_i | \His^C_t\}} \bracket{I_t}\\
    &\stackrel{\text{(b)}}{=} \beta \frac{S_t}{N} \E_{\{t^R_i | \His^C_t\}} \bracket{\sum_{t^I_{i} \in \His^C_t} \mathds{1}(t^R_{i} - t^I_{i} > t - t^I_{i})} \\
    &\stackrel{\text{(c)}}{=} \beta \frac{S_t}{N} \sum_{t^I_{i} \in \His^C_t} \E_{\{t^R_i | \His^C_t\}} \bracket{\mathds{1}(t^R_{i} - t^I_{i} > t - t^I_{i})} \\
    &\stackrel{\text{(d)}}{=} \beta \frac{S_t}{N}\sum_{t^I_{i} \in \His^C_t}  \int^\infty_0 \mathds{1}(\tau_i > t - t^I_{i}) f(\tau_i | \His^C_t) d\tau_i \\
    &\stackrel{\text{(e)}}{=}  \beta \frac{S_t}{N} \sum_{t^I_{i} \in \His^C_t} \int^\infty_{t - t^I_{i}}f(\tau_i | \His^C_t) d\tau_i \\
    &\stackrel{\text{(f)}}{\geq}  \beta \frac{S_t}{N} \sum_{t^I_{i} \in \His^C_t} \int^\infty_{t - t^I_{i}}f(\tau_i) d\tau_i \numberthis \label{eq:derivation2}
\end{align*}
Here Eq.~(\ref{eq:derivation2}a) is due to the independence between $S_t$ and $t^R_i$ given $\His^C_t$. 
Eq.~(\ref{eq:derivation2}b) follows from decomposing the step-wise stochastic process $I_t$ --- that the recovery time of each individual therein is greater than time $t$, i.e., $t_i^R > t$. By definition $t_i^R > t_i^I$, we can subtract infection times $t_i^I$ on both sides and preserve the sign of the inequality --- leading to $t_i^R - t_i^I > t - t_i^I$, which is easier to model since the left hand side correspond to the recovery time of the $i$-th infection.
$\mathds{1}(x)$ is an indicator function that takes value $1$ if the proposition $x$ is true, $0$ otherwise. Eq.~(\ref{eq:derivation2}c) pushes the expectation into the summation due to known infection events.
Eq.~(\ref{eq:derivation2}d) expands the expectation for each recovery time, and uses $\tau_i = t^R_i - t^I_i \sim f(\tau_i | \His^C_t)$ where the $i^{th}$ individual's recovery time distribution is conditional.
Eq.~(\ref{eq:derivation2}e) uses the definition of the indicator function to change the lower bound of integration.

To show the inequality in Eq.~(\ref{eq:derivation2}f), we reduce it down to proofing
\begin{align}
    \int^\infty_{t - t^I_{i}}f(\tau_i | \His^C_t) d\tau_i \geq \int^\infty_{t - t^I_{i}}f(\tau_i) d\tau_i \Longrightarrow
    \int^{t - t^I_{i}}_0 f(\tau_i | \His^C_t) d\tau_i \leq \int^{t - t^I_{i}}_0 f(\tau_i) d\tau_i
\end{align}
which is equivalent to $\Prob \bracket{t^R_i < t | \His^C_t} \leq \Prob \bracket{t^R_i < t}$. To proof this, we reason it from the perspective of a branching structure, namely any future event is a descendent event triggered by past event. We denote the infection events triggered by $i^{th}$ individual as $\His^C_i = \{t^I_j | t^I_j \in \His^C_t, t^I_i < t^I_j < t\}$ and use $t^I_m = \max\{\His^C_i\}$. Then we can see that $\Prob \bracket{t^R_i < t | \His^C_t} = \Prob \bracket{t^R_i < t | \His^C_i}$. Two possible cases emerge
\begin{itemize}
    \item $\His^C_i = \emptyset$: we have $\Prob \bracket{t^R_i < t | \His^C_i} = \Prob \bracket{t^R_i < t}$ where the equality holds.
    \item $|\His^C_i| > 0$: we can also reduce the dependency with $\Prob \bracket{t^R_i < t | \His^C_i} = \Prob \bracket{t^I_m < t^R_i < t}$. We then compare $\Prob \bracket{t^I_m < t^R_i < t} = \int_{t^I_m - t^I_i}^{t-t^I_i} f(\tau) d\tau$ against $\Prob \bracket{t^R_i < t} = \int_{0}^{t-t^I_i} f(\tau) \tau$ given recovery time distribution $f(t)$. As $f(t) \geq 0, \forall t \in R$, we conclude $\int_{0}^{t-t^I_i} f(\tau) \tau > \int_{t^I_m - t^I_i}^{t-t^I_i} f(\tau) d\tau$
\end{itemize}
Overall, that proofs $\int^\infty_{t - t^I_{i}}f(\tau_i | \His^C_t) d\tau_i \geq \int^\infty_{t - t^I_{i}}f(\tau_i) d\tau_i$.

\subsection{Emprical analysis of the intensity difference}
\begin{figure*}[!tpb]
    \begin{subfigure}{0.45\textwidth}
        \includegraphics[width=\textwidth,page=1]{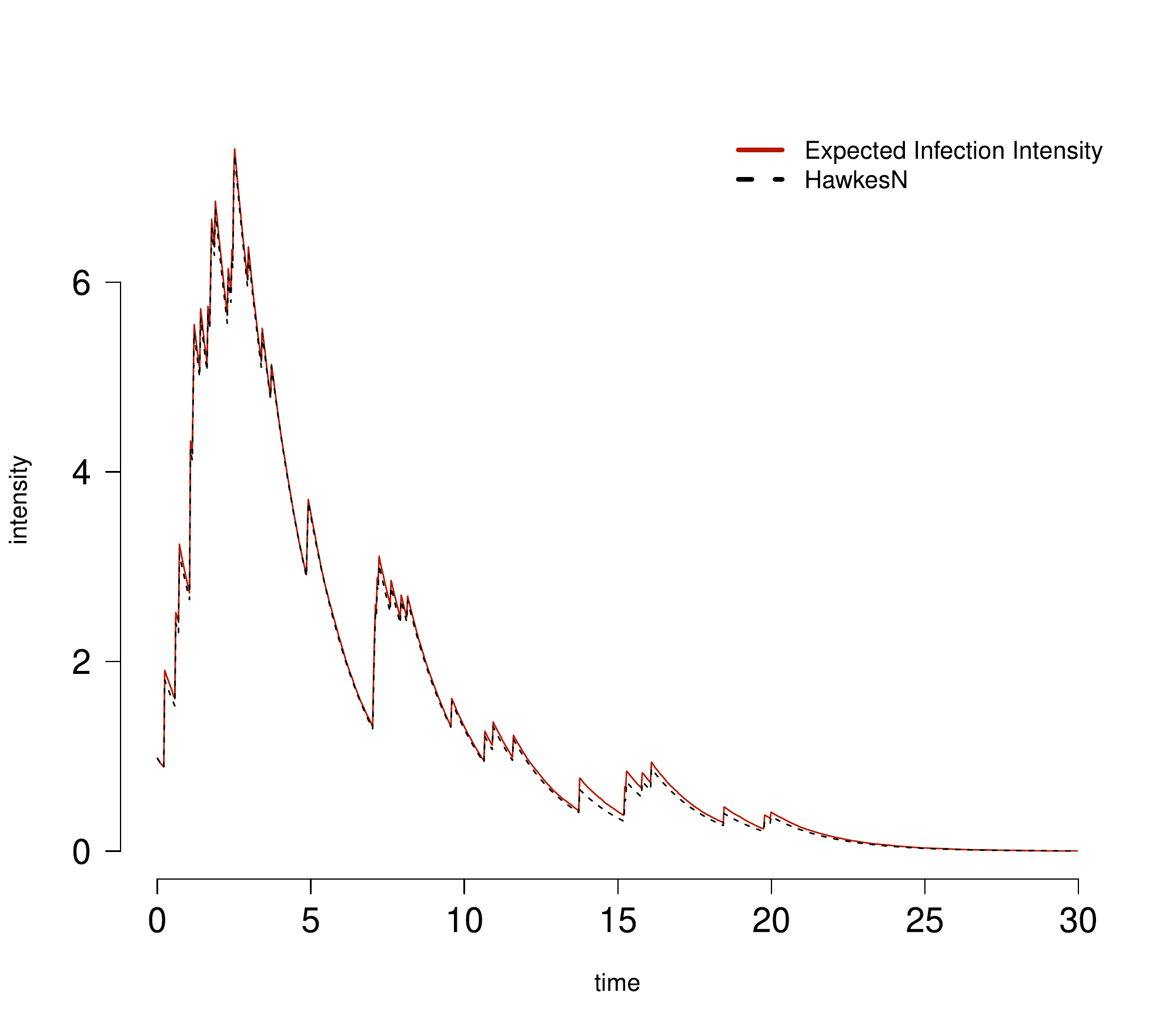}
        \caption{Intensity difference between the expected infection intensity and HawkesN intensity (an exponential recovery time distribution: $N=50, \beta = 1, \gamma = 0.5$). The expected infection intensity is approximated with $2000$ simulations.}
    \label{subfig:intensity-difference-example}
    \end{subfigure}
    \hfill
    \begin{subfigure}{0.38\textwidth}
        \includegraphics[width=\textwidth,page=1]{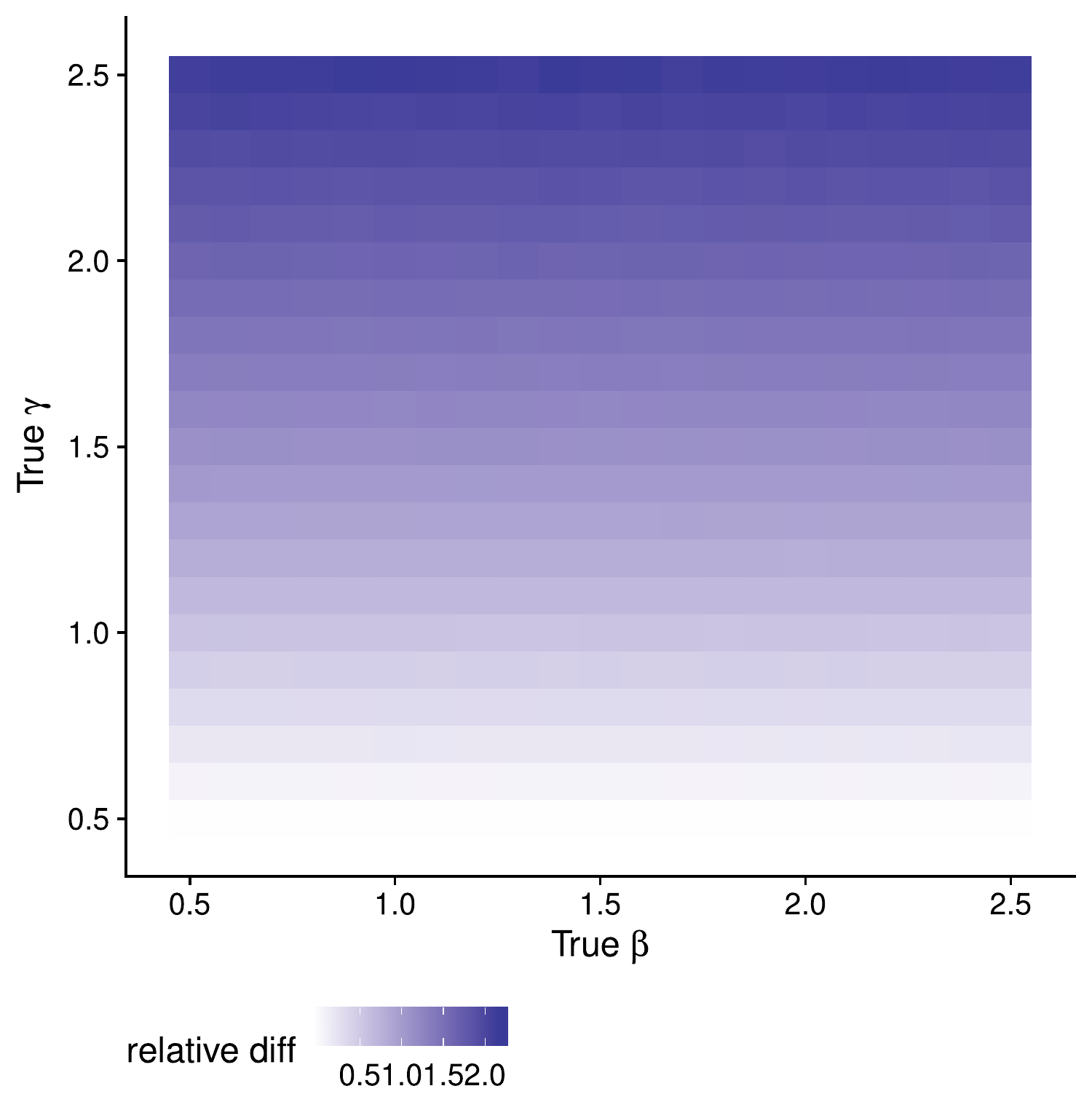}
        \caption{Given the infection events in (a), the relative difference between the expected infection intensity values and the HawkesN intensity values given various parameters.}
        \label{subfig:intensity-difference-more}
    \end{subfigure}
    \caption{}
    \label{fig:intensity-difference}
\end{figure*}

In this section, we explore the difference between the expected infection intensity values of generalized stochastic SIR and the corresponding HawkesN intensity. \cref{subfig:intensity-difference-example} presents an example of this intensity difference given a specific parameter set. For a given stochastic SIR process, we approximate its expected infection intensity by fixing the infection events and simulating recovery events via standard rejection-sampling technique~\citepAP{Daley2008}. HawkesN intensity values, on the other hand, can be computed from~\cref{eq:hawkesn}. \cref{subfig:intensity-difference-more} explores the relative intensity difference between the two at various parameter combinations.

\subsection{Branching factor}
For the classic SIR, the reproduction number $R_0$ is defined as \citepAP{newman2018networks} 
\begin{equation}
    R_0 =  \int^{\infty}_0 \beta \tau \gamma e^{-\gamma \tau} d\tau
\end{equation}
where $\beta \tau$ is the expected number of individuals contacted by an infected individual and integrating it with the recovery time distribution leads to $R_0$. We can then express it with a general recovery time distribution as $R_0 =  \int^{\infty}_0 \beta \tau f(\tau) d\tau$. We show its equivalence to~\cref{eq:branching_factor} as following
\begin{align*}
    R_0 &=  \int^{\infty}_0 \beta \tau f(\tau) d\tau \\
    &= \beta \bracket{\tau \int^\tau_0 f(\eta) d\eta}^{\infty}_0 - \beta \int^{\infty}_0 \int^{\tau}_0 f(\eta) d\eta d\tau \tag{Integration by parts}\\
    &= \beta \bracket{\tau \int^\tau_0 f(\eta) d\eta}^{\infty}_0 - \beta \int^{\infty}_0 (1 - \int^{\infty}_{\tau} f(\eta)) d\eta d\tau \tag{due to $\int^{\infty}_0 f(\eta) d\eta = 1$}\\
    &= \beta \bracket{\tau \int^\tau_0 f(\eta) d\eta}^{\infty}_0 - \beta \bracket{\tau}^{\infty}_0 + \beta \int^{\infty}_0 \int^{\infty}_{\tau} f(\eta) d\eta d\tau \\
    &= \beta \bracket{\tau \int^\tau_0 f(\eta) d\eta}_{\tau = \infty} - \beta \bracket{\tau}_{\tau = \infty} + \beta \int^{\infty}_0 \int^{\infty}_{\tau} f(\eta) d\eta d\tau  \tag{due to $\int^{0}_0 f(\eta) d\eta = 0$}\\
    &=\beta \int^{\infty}_0 \int^{\infty}_{\tau} f(\eta) d\eta d\tau \tag{due to $\int^{\infty}_0 f(\eta) d\eta = 1$}
\end{align*}

\section{Likelihood and Parameter Estimation}
We conduct maximum likelihood estimation for parameter inference. For total population size $N$, we adopt the practice from both \citetAP{jin2013epidemiological} and \citetAP{Rizoiu2017c}: fitting $N$ as an unknown parameter. Let $\Theta^E$ denote the set of all parameters in the stochastic SIR models, e.g., $\Theta^E = \{\beta, \gamma, N\}$ for stochastic SIR described in Section~\ref{sec:preliminaries}. To estimate $\Theta^E$ from a given stochastic SIR process until time $t$ ($\His^C_t, \His^R_t$) with a recovery time distribution $f(t)$, the likelihood function of $\Theta^E$ can be expressed based on the log-likelihood estimator for point processes \citepAP{Daley2008} as
\begin{align}\label{eq:likelihood1}
    \mathcal{L}(\Theta^E; \His^C_t, \His^R_t) =& \sum_{t^I_i \in \His^C_t} \log \lambda^I (t^I_i) - \int_0^{t} \lambda^I (\tau) d\tau  + \sum_{i \in U(\His^R_t)} \log f(t^R_i - t^I_i) \numberthis
\end{align}
The first two terms of RHS of \cref{eq:likelihood1} comes from
\begin{align}\label{eq:likelihood2}
    \mathcal{L}(\Theta^E; \His^C_t, \His^R_t) &= \log \prod_{t^I_i \in \His^C_t} \lambda^I(t^I_i) e^{-\int_0^{t^I_i} \lambda^I(u) du} = \log \prod_{t^I_i \in \His^C_t} \lambda^I(t^I_i) e^{-\int_{t^I_{i-1}}^{t^I_i} \lambda^I(u) du} \\
    &= \log e^{-\int_{0}^{max\{\His^C_t\}} \lambda^I(u) du} \prod_{t^I_i \in \His^C_t} \lambda^I(t^I_i) \\
    &=- \int_0^{t} \lambda^I (\tau) d\tau + \sum_{t^I_i \in \His^C_t} \log \lambda^I (t^I_i) 
\end{align}

When recovery events are not observed, i.e., only $\His^C$ is presented, we take expectation over the recovery event history $\His^R$ on \cref{eq:likelihood1}:
HawkesN log-likelihood functions after the solving integral part in \cref{eq:likelihood2} with different kernel functions are listed as following:
\begin{itemize}
    \item Exponential
    \begin{align*}
        \mathcal{L}_{EXP}(\Theta^H; \His^C)  = \sum_{j=2}^n \log\left(\lambda^H\left(t_j^-\right)\right) - \kappa \sum_{j=1}^{n-1 } \sum_{l=j}^{n-1} \frac{N - l}{N} \left[e^{-\theta(t_l - t_j)} - e^{-\theta(t_{l+1} - t_j)} \right]
    \end{align*}
    
    \item Power-law
    \begin{align*}
        \mathcal{L}_{PL}(\Theta^H; \His^C) =& \sum_{t_i \in \His^C} \log \frac{N - i}{N} \kappa \sum_{t_j \in \His^C, t_j < t_i} (t_i - t_j + c)^{-(1+\theta)}  \\
        &- \kappa \theta \sum_{t_i \in \His^C, t_i < t_n} \sum_{t_j \in \His^C, t_i \leq t_j < t_n} \frac{N - j}{N} \bracket{(t_j - t_i + c)^{-\theta} - (t_{j+1} - t_i + c)^{-\theta}}
    \end{align*}
    
    \item Tsallis Q-EXP
    \begin{align*}
        \mathcal{L}_{Q-EXP}(\Theta^H; \His^C) =& \sum_{t_i \in \His^C} \log \frac{N - i}{N} \kappa \sum_{t_j \in \His^C, t_j < t_i} \bracket{1 + (\theta - 1) (t_i - t_j)}^{\frac{1}{1 - \theta}} \\
        &\hspace{-2cm}- \frac{\kappa}{2 - \theta} \sum_{t_i \in \His^C, t_i < t_n} \sum_{t_j \in \His^C, t_i \leq t_j < t_n}\frac{N - j}{N} \bracket{\bracket{1 + (\theta - 1)(t_j - t_i)}^{\frac{2-\theta}{1-\theta}} - \bracket{1 + (\theta - 1)(t_{j+1} - t_i)}^{\frac{2 - \theta}{1 - \theta}}}
    \end{align*}
\end{itemize}

Some natural constraints are applied on $N \geq C_t$ and on other parameters as in~\cref{tab:decay-function-equivalence}. Eqs. (\ref{eq:likelihood1}) (\ref{eq:likelihood2}) are non-linear functions and we use an optimization tool AMPL~\citepAP{fourer1987ampl} bridged with a non-linear solver Ipopt~\citepAP{Wachter2006} for maximizing them and estimating model parameters.

After obtaining $\Theta^H$, \cref{eq:equivalence-f-phi} leads us to corresponding stochastic SIR parameters $\Theta^E$. Similarly, \cref{eq:f-to-phi} links $\Theta^E$ inferred by \cref{eq:likelihood1} to $\Theta^H$. This helps one reveal the underlying recovery processes when missing recovery event data or concentrate on infection process yet leveraging both infection and recovery events in data.

\section{Extra experiment results}
\begin{figure*}[!tpb]
    \centering
    \begin{subfigure}{0.45\textwidth}
        \includegraphics[width=\textwidth,page=1]{images/{best-fit-model-analysis-News}.pdf}
        \caption{\textit{NEWS}: EXPN vs. PL }
    \end{subfigure}
    \begin{subfigure}{0.45\textwidth}
        \includegraphics[width=\textwidth,page=1]{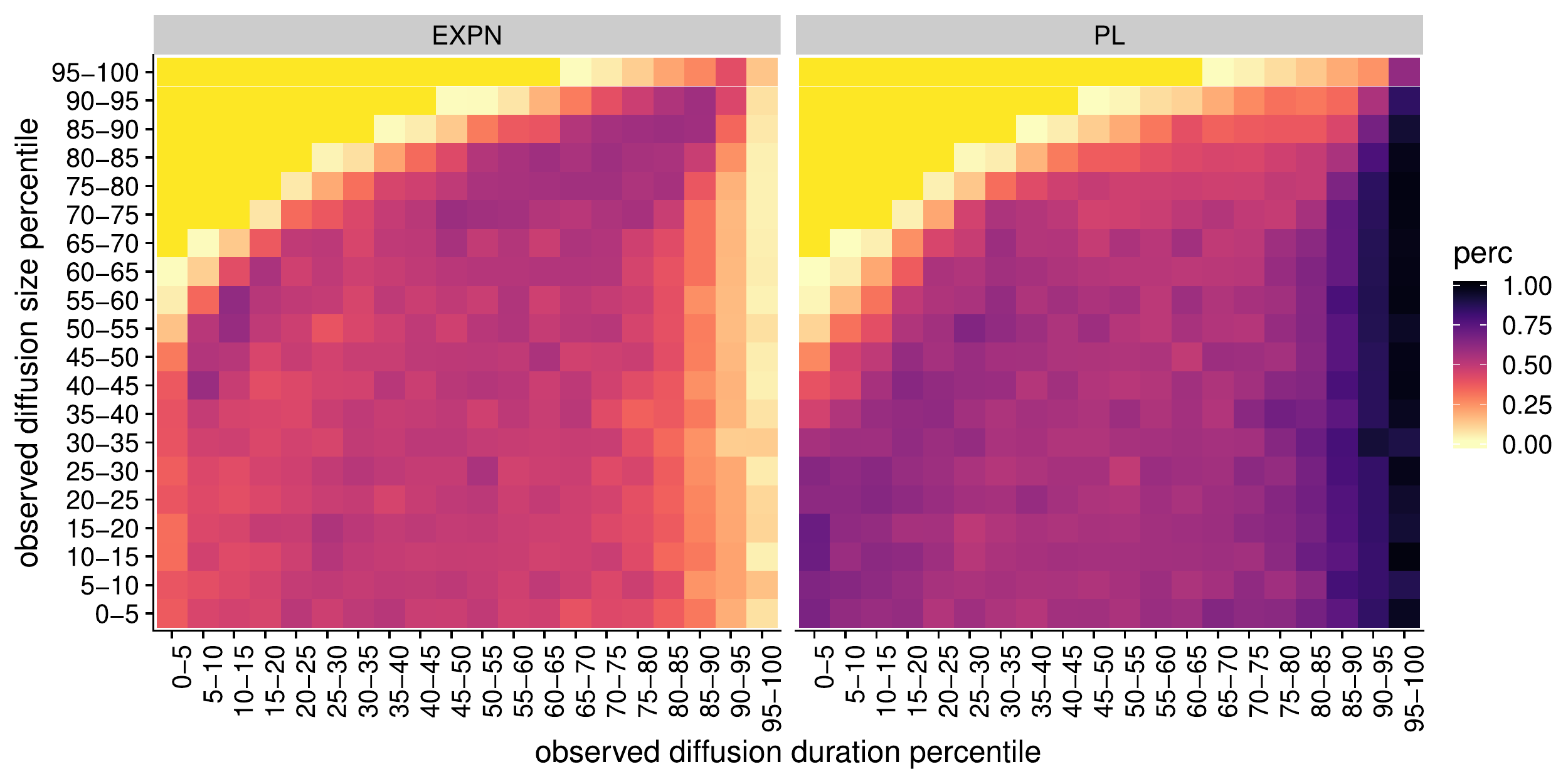}
        \caption{\textit{Seismic}: EXPN vs. PL }
    \end{subfigure}
    \begin{subfigure}{0.45\textwidth}
        \includegraphics[width=\textwidth,page=1]{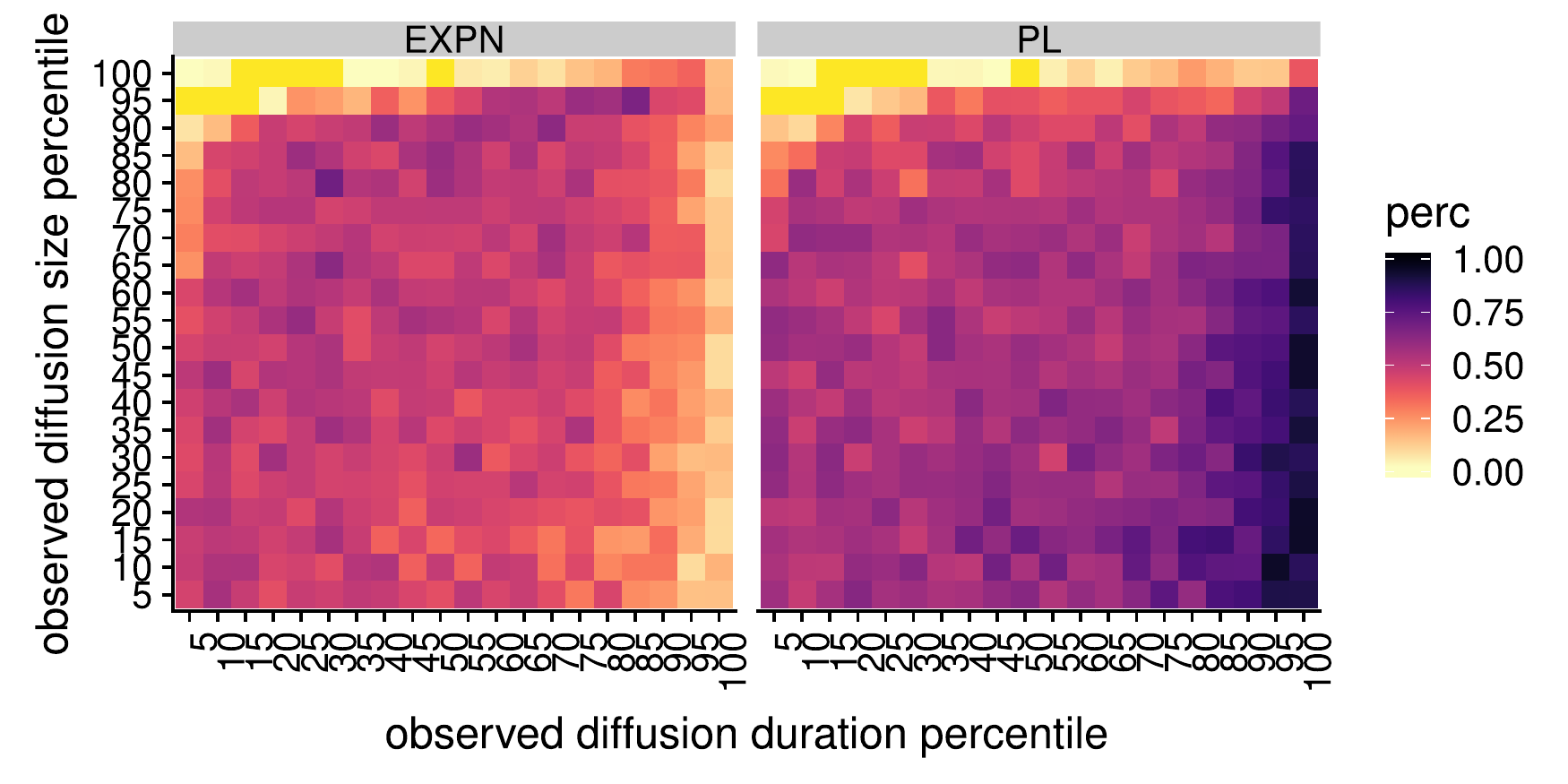}
        \caption{\textit{ActiveRT}: EXPN vs. PL }
    \end{subfigure}
    \caption{}
    \label{fig:all-model-comparison}
\end{figure*}

\begin{figure*}[!tpb]
    \centering
    \begin{subfigure}{0.33\textwidth}
        \includegraphics[width=\textwidth,page=2]{images/{2017-07-07-holdout-nll-perc-news-0.40-unmarked}.pdf}
        \caption{\textit{NEWS} }
    \end{subfigure}
    \begin{subfigure}{0.33\textwidth}
        \includegraphics[width=\textwidth,page=2]{images/{2017-07-07-holdout-nll-perc-activeyt-0.40-unmarked}.pdf}
        \caption{\textit{ActiveRT} }
    \end{subfigure}
    \begin{subfigure}{0.33\textwidth}
        \includegraphics[width=\textwidth,page=2]{images/{2017-07-07-holdout-nll-perc-seismic-0.40-unmarked}.pdf}
        \caption{\textit{Seismic} }
    \end{subfigure}
    \begin{subfigure}{0.33\textwidth}
        \includegraphics[width=\textwidth,page=2]{images/{2017-07-07-holdout-nll-perc-news-0.40}.pdf}
        \caption{\textit{NEWS} with marks}
    \end{subfigure}
    \begin{subfigure}{0.33\textwidth}
        \includegraphics[width=\textwidth,page=2]{images/{2017-07-07-holdout-nll-perc-activeyt-0.40}.pdf}
        \caption{\textit{ActiveRT} with marks}
    \end{subfigure}
    \begin{subfigure}{0.33\textwidth}
        \includegraphics[width=\textwidth,page=2]{images/{2017-07-07-holdout-nll-perc-seismic-0.40}.pdf}
        \caption{\textit{Seismic} with marks}
    \end{subfigure}

    \begin{subfigure}{0.45\textwidth}
        \includegraphics[width=\textwidth]{images/{popularity-prediction-News-3600}.pdf}
        \caption{Popularity prediction on \textit{News}}
        \label{subfig:prediction-news-2}
        \vspace{-3mm}
    \end{subfigure}
    \begin{subfigure}{0.45\textwidth}
        \includegraphics[width=\textwidth]{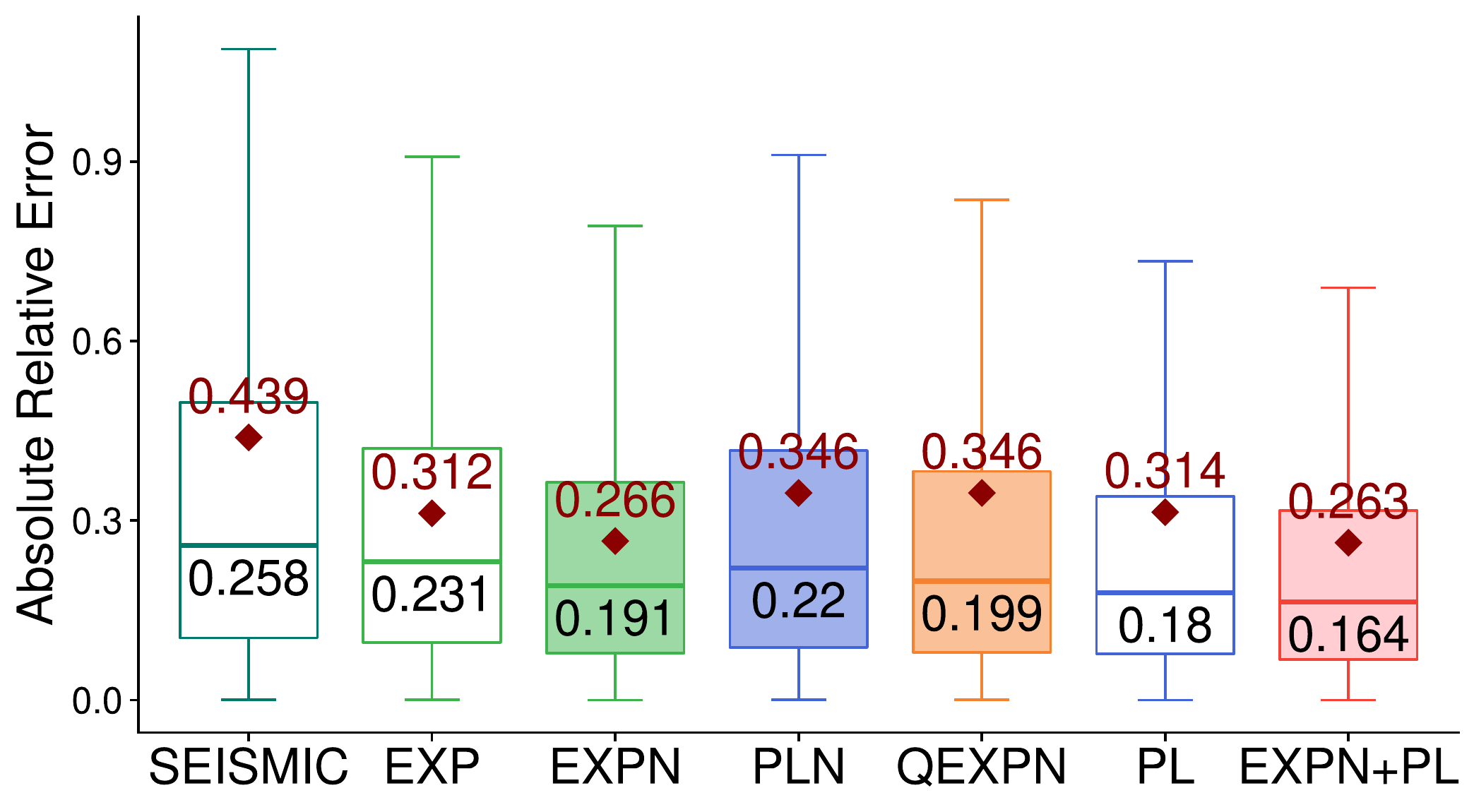}
        \caption{Popularity prediction on \textit{ActiveRT}}
        \label{subfig:prediction-activeyt}
        \vspace{-3mm}
    \end{subfigure}
    \begin{subfigure}{0.45\textwidth}
        \includegraphics[width=\textwidth]{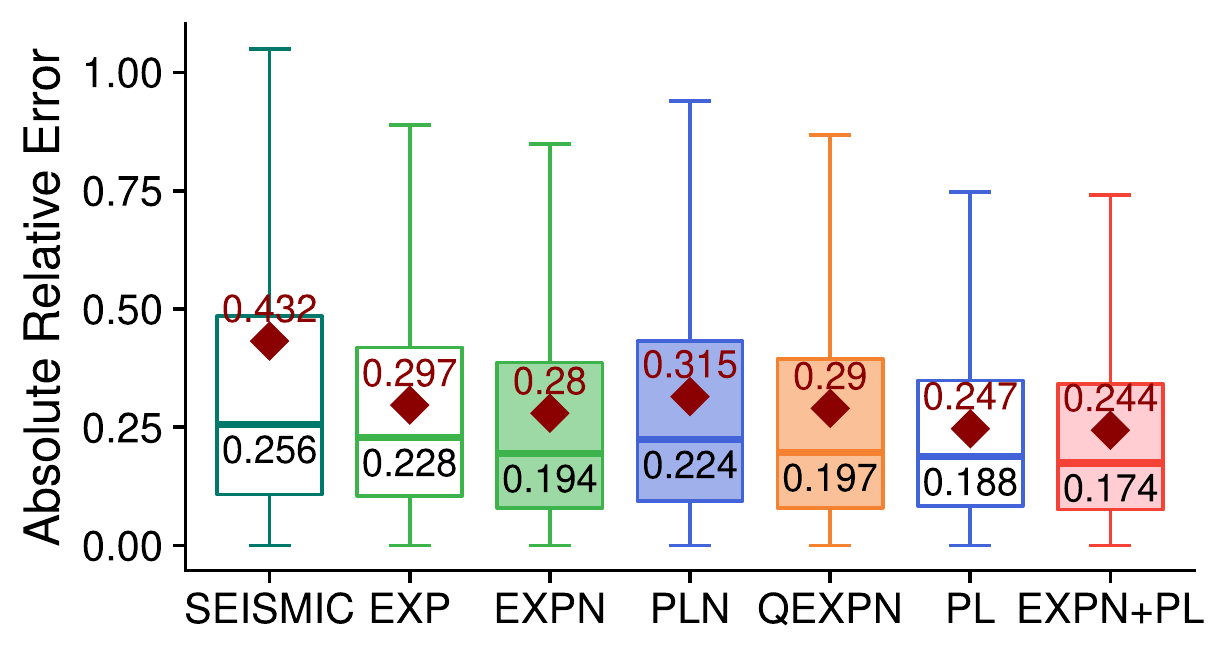}
        \caption{Popularity prediction on \textit{Seismic}}
        \label{subfig:prediction-seismic}
        \vspace{-3mm}
    \end{subfigure}
    \caption{
        Fig.~\textit{(a)-(f)} are holdout negative log-likelihood
        of models on three datasets. Fig.~\textit{(g)-(i)} are popularity prediction performance on three datasets
    }
	\label{fig:holdout-experiments-marks}
\end{figure*}

\cref{fig:all-model-comparison} presents the comparison of a most distinct model pair on each dataset.
\cref{fig:holdout-experiments-marks} shows the holdout log-likelihood values and popularity prediction performance of models fitted with event marks on all three retweet cascade datasets.

\section{Compare size distributions of stochastic SIR and HawkesN}
\begin{figure}[!tbp]
	\centering
    \begin{subfigure}{0.6\textwidth}
        \includegraphics[width=\textwidth,page=1]{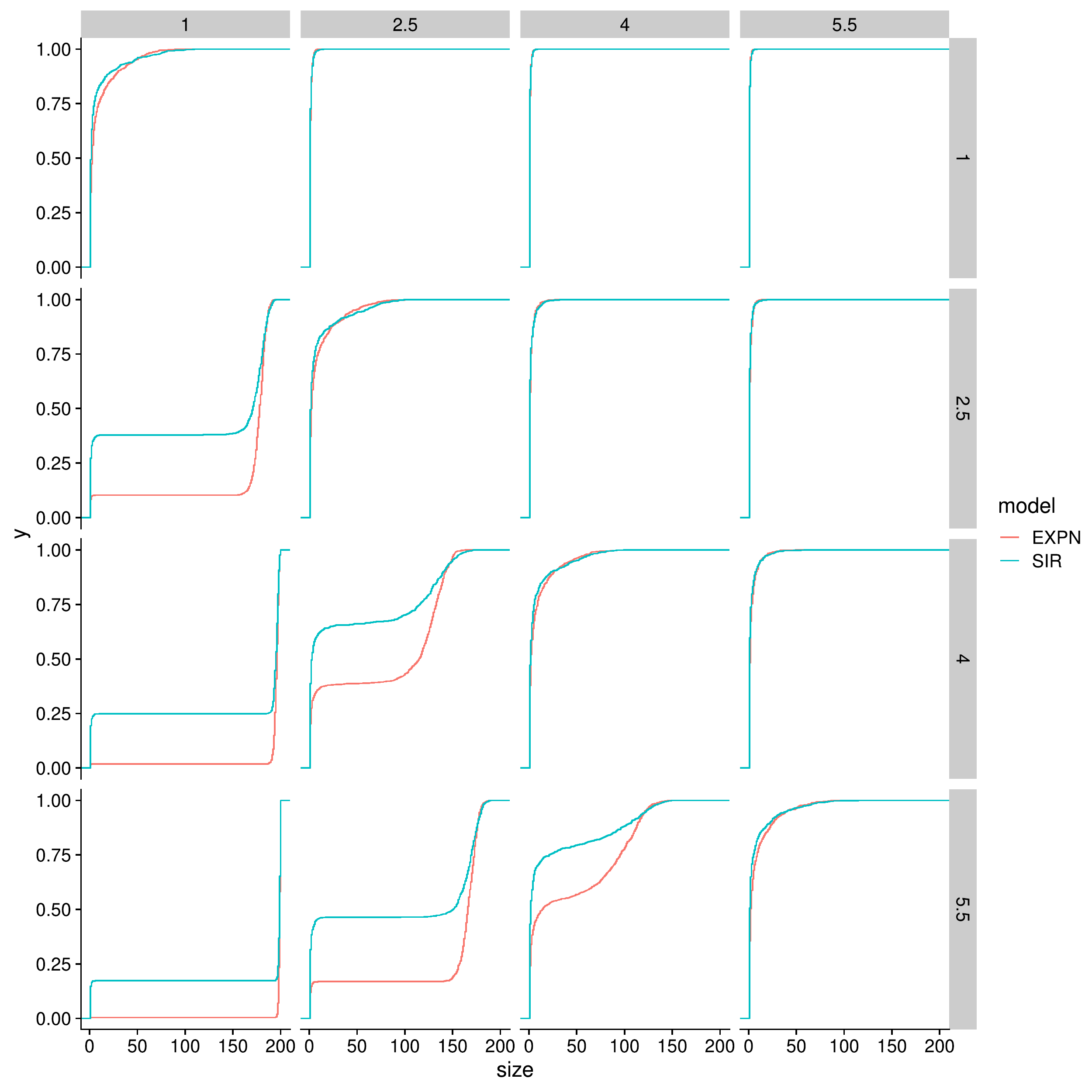}
        \caption{EXPN vs stochastic SIR with an exponential recovery distribution}
        \label{subfig:size-dist-exp}
    \end{subfigure}
    \begin{subfigure}{0.6\textwidth}
        \includegraphics[width=\textwidth,page=2]{images/2019-10-24-same-param-hawkes-vs-sir}
        \caption{PLN vs stochastic SIR with a power-law recovery distribution}
        \label{subfig:size-dist-pl}
    \end{subfigure}
    \caption{Size distribution comparison between HawkesN models and stochastic SIR processes. The x axis and y axis are different parameter values ($\beta,\gamma$ for (a) and $c, \theta$ for (b), respectively). For each small plot, the x axises are the cascade sizes while the y axises are the corresponding empirical cumulative densities.
	}
	\label{fig:size-dist}
\end{figure}

In this section, we empirically show the difference of stochastic SIR and HawkesN in terms of their process final size distributions given same set of parameters. We approximate their final size distributions (empirical cumulative density functions) via simulation with $1000$ simulated processes for each model and parameter set. \cref{fig:size-dist} shows the results given different sets of parameters. From this plot, we notice that stochastic SIR processes consistently present higher probabilities of smaller cascade sizes and HawkesN models tend to generate larger size cascades for high branching factors.

\begin{figure}[!tbp]
	\centering
    \includegraphics[width=0.4\textwidth]{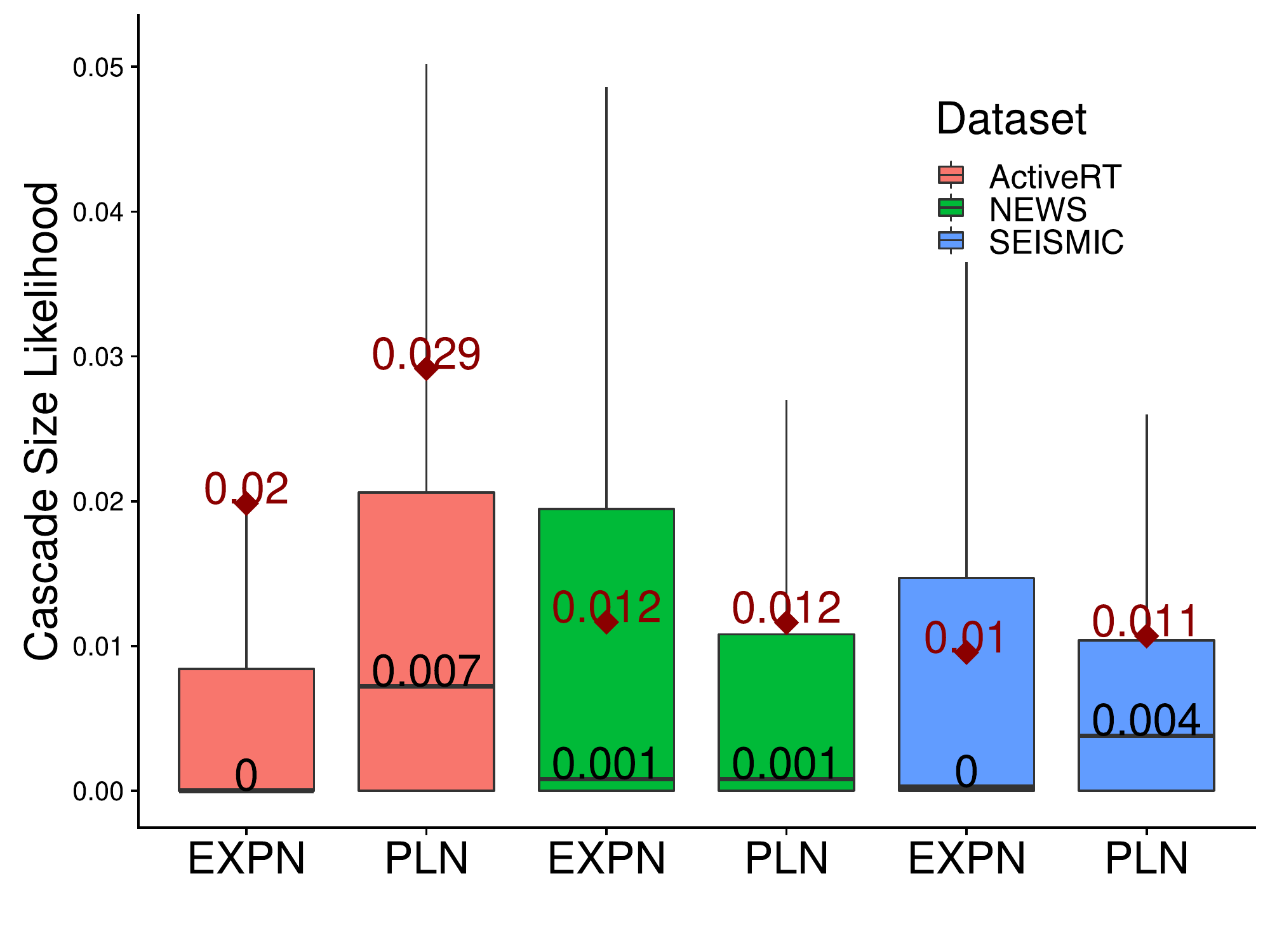}
	\caption{
	    The likelihood of observed cascade sizes on \textit{ActiveRT}, \textit{NEWS} and \textit{Seismic}, using EXPN and PLN.
	    The parameters of each model are fitted using $40\%$ of the events in each cascade. 
	    The distributions of cascade size are approximated using 5000 simulations for each set of parameters --- higher is better.
    }
	\label{fig:power-law-diffusion-size}
\end{figure}

\section{Likelihood of cascade sizes.}
We study the probability distribution of final size for Twitter cascades.
\citetAP{Rizoiu2017c} have proposed a Markov chain method to estimate the distribution of final cascade size, based on SIR's memory-less property.
However, stochastic SIR with non-exponentially distributed recovery times do not have this property.
To overcome this problem, we employ a simulation-based computation of the size distribution.
Given a parameter set $\Theta^E$ --- the parameters of HawkesN fitted on $k$ events ---
we approximate the size distributions $\Prob \left[ |\His^C| = j \middle| \Theta^{E}, j \geq k \right]$ 
by converting the HawkesN parameters to stochastic SIR parameters, and applying \cref{alg:simulation} to simulate $5000$ realizations for each cascade.
We construct the empirical size distribution by aggregating the sizes of the realizations and smoothing the obtained distribution.
Given $n$ the observed final size of a cascade, its likelihood under the constructed distribution is
$\Prob \left[ |\His^C| = n \middle| \Theta^{E} \right]$. 

We employ the above methodology to compute the likelihood of the observed final size for three samples --- one for each dataset --- each sample containing $1000$ cascades.
For every cascade we construct two size distributions, using HawkesN with an exponential and a power-law kernel respectively.
Figure~\ref{fig:power-law-diffusion-size} aggregates the computed likelihoods, per dataset and per HawkesN kernel type.
Each boxplot contains $1000$ datapoints.
We observe that for \textit{ActiveRT} and for \textit{Seismic}, the observed final size is more likely under the distribution constructed using power-law kernel than under the exponential kernel.
This is likely due to power-law kernel being long-tailed, 
and able to explain the minority of very large cascades occurring naturally~\citepAP{Goel2015}.
For \textit{NEWS}, the similar performances of the two kernels 
are likely due to news being time-sensitive content, and on average having smaller cascades.

\bibliographystyleAP{ACM-Reference-Format}
\bibliographyAP{acm}
\end{document}